\documentclass{JHEP}
\usepackage{graphicx}
\usepackage{epsfig}
\usepackage{amssymb}
\usepackage{amsmath}
\usepackage{float}
\usepackage{subfigure, amsmath,graphicx}
\usepackage{cite}

\newcommand{\bea}{\begin{eqnarray}}
\newcommand{\eea}{\end{eqnarray}}
\newcommand{\bean}{\begin{eqnarray*}}
\newcommand{\eean}{\end{eqnarray*}}
\newcommand{\nn}{\nonumber \\}

\def\O #1{\overline{#1}}

\def\W #1{\widetilde{#1}}
\def\WH #1{\widehat{#1}}

\def\eref#1{(\ref{#1})}

\def\d{{\rm d}}

\def\b{{\beta}}

\def\d{\partial}

\def\la{\lambda}
\def\eps{\epsilon}

\def\Spbb #1{ \left[ #1 \right]}
\def\Spaa #1{ \left\langle #1 \right\rangle}
\def\Spab #1{ \left\langle #1 \right]}

\def\ket#1{\left| #1\right\rangle}

\def\bket#1{\left| #1\right]}

\def\Label#1{\label{#1}%
  \smash{\hbox to0pt{\raise1ex\hbox{\tiny[#1]}\hss}}}

\title{Integral Reduction by Unitarity Method for Two-loop Amplitudes: A Case Study }

\author{Bo Feng $^{a,b}$, Jun
Zhen $^{a}$, Rijun Huang $^{c}$, Kang Zhou
$^{a}$\footnote{The unusual ordering of authors is just to satisfy
outdated requirement for Ph. Degree Of Zhejiang University in
China.}\\$^a$Zhejiang Institute of Modern Physics, Zhejiang
University, Hangzhou, 310027, P. R. China\\$^b$Center of
Mathematical Sciences, Zhejiang University, Hangzhou, 310027, P. R.
China\\$^c$Institut de Physique Th\'eorique, CEA-Saclay, F--91191
Gif-sur-Yvette cedex, France}

\date{\today}
\abstract{In this paper, we generalize the unitarity method to two-loop diagrams and use it to discuss the integral bases of reduction.
To test out method, we focus on the four-point double-box diagram as
well as its related daughter diagrams, i.e., the double-triangle
diagram and the triangle-box diagram. For later two kinds of diagrams, we
have given complete analytical results in general
$(4-2\eps)$-dimension.}


\begin{document}

\section{Introduction}
\setcounter{equation}{0}

Currently, the focus of high energy physics is the LHC experiment.
To understand the experiment data, we need to evaluate scattering
amplitudes to high accuracy level required by data. Thus for most
processes, the one-loop evaluation becomes necessary. In last ten
years, enormous progress has been made in the computation of
one-loop scattering amplitudes(see, for example, the references
\cite{Bern:2008ef,Binoth:2010ra,AlcarazMaestre:2012vp} and citations
in the papers). However, for some processes in modern colliders,
such as the process $gg\to \gamma\gamma$ which is an important
background for searching the Higgs boson at the LHC, one-loop
amplitudes do not suffice since their leading-order terms begin at
one loop. Thus next-to-leading order corrections  require the
computation of two-loop amplitudes
\cite{Berger:1983yi,Aurenche:1985yk,Ellis:1987xu}.

The traditional method for amplitude calculation is through the
Feynman diagram. This method is well organized and has clear
physical picture. It has also been implemented into many computer
programs. However, with increasing of loop level or the number of
external particles, the complexity of computation increases
dramatically. Thus even with the most powerful computer available,
many interesting processes related to LHC experiments can not be
dealt by the traditional method.

To solve the challenge, many new methods(see books
\cite{Smirnov:2004ym,Smirnov:2006ry,Smirnov:2012gma}) have been
developed, such as IBP(integrate-by-part) method
\cite{Chetyrkin:1981qh,Tarasov:1998nx,Bern:2000dn,Anastasiou:2000kg,Anastasiou:2000ue,Glover:2001af,
Anastasiou:2001sv,Laporta:2001dd,Bern:2002tk,Tarasov:2004ks}(some
new developments, see
\cite{Gluza:2010ws,Kalmykov:2011yy,Schabinger:2011dz}), differential
equation method
\cite{Kotikov:1990kg,Remiddi:1997ny,Gehrmann:1999as,Argeri:2007up,Henn:2013pwa,Henn:2013woa,
Henn:2013nsa,Argeri:2014qva}, MB(Mellin-Barnes) method
\cite{Bergere:1973fq,Usyukina:1975yg,Smirnov:1999gc,Tausk:1999vh},
etc. Among these methods, the reduction method
\cite{Passarino:1978jh,Neerven:1984,Bern:1992,Ellis:2007} is one of
the most useful methods. More explicitly, the reduction of an
amplitude means that any amplitude ${\cal A}$ can be expanded by
bases(or "master integral") as
\bea {\cal A}=\sum_ic_i {\cal A}_i~,~~~\label{1loop-exp} \eea
with rational coefficients $c_i$. With this expansion, the amplitude
calculation can be separated into two parts: (a) the evaluation of
bases(or master integrals) at given loop order and (b) the
determination of coefficients $c_i$ for a particular process. For
the former part, it can be done once for all and the results can be
applied to any process. Thus in the practical application, the
latter part, i.e., the determination of coefficients, becomes the
central focus of all calculations.

Unitarity method is an ideal tool to determine
coefficients\cite{Bern:1994zx,Bern:1995db,Bern:1997sc,Bern:1996ja,Britto:2004nc,Britto:2004nj,
Bern:2005cq,Britto:2005ha,Brandhuber:2005jw,Bern:2007dw,Forde:2007mi,
Badger:2008cm,Anastasiou:2006jv,Britto:2006fc,Berger:2009zb,Bern:2010qa}.
With the expansion \eref{1loop-exp}, if we perform unitarity cut on
both sides, we will get
\bea \Delta{\cal A}=\sum_ic_i\Delta {\cal
A}_i~.~~~\label{1loop-exp-1} \eea
So if both $\Delta{\cal A}$ and $\Delta {\cal A}_i$ can be evaluated
analytically, and if different $\Delta {\cal A}_i$ has
distinguishable analytic structure(which we will call the
"signature" of basis under the unitarity cut), we can compare both
sides of \eref{1loop-exp-1} to determine coefficients $c_i$,
analogous to the fact that if two polynomials of $x$ are equal, so
are their coefficients of each term $x^n$. The unitarity method has
been proven to be very successful in determining coefficients for
one-loop amplitudes(see reviews \cite{Britto:2010xq,
Dixon:2013uaa}). For some subsets of bases(such as box topology for
one-loop and double-box topology for planar two-loop), more
efficient method, the so called "generalized unitarity method"(or
"maximum unitarity cut" or "leading singularity"), has been
developed
\cite{Britto:2004nc,Buchbinder:2005wp,ArkaniHamed:2009dn,ArkaniHamed:2012nw,Kosower:2011ty,Larsen:2012sx,CaronHuot:2012ab,Johansson:2012zv,
Johansson:2012sf,Sogaard:2013yga,Johansson:2013sda,Sogaard:2013fpa}.

The applicability of reduction method is based on the valid
expression of expansion \eref{1loop-exp}. Thus the determination of
bases becomes the first issue. From recent study, it is realized
that there are two kinds of bases: the integrand bases and the
integral bases. The integrand bases are algebraically independent
rational functions before performing loop integration. For one-loop,
the integrand bases have been determined by OPP\cite{Ossola:2006us}.
For two-loop or more, the computational algebraic geometry method
has been proposed to determine the integrand bases
\cite{Mastrolia:2011pr,Badger:2012dp,Mastrolia:2012an,Kleiss:2012yv,Badger:2012dv,
Mastrolia:2012wf,Huang:2013kh,Badger:2013gxa,Zhang:2012ce,Feng:2012bm,Mastrolia:2012du}.

In general the number of integrand bases is larger than the number
of integral bases, because after loop integration, some combinations
of integrand bases may vanish. For one-loop amplitudes, the
difference between these two numbers  is not very significant. For
example, the number of integral bases is one while the number of
integrand bases is seven for triangle topology of renormalizable
field theories\cite{Ossola:2006us}. However, for two-loop
amplitudes, the difference could be huge. As we will show later, for
double-triangle topology, there are only several integral bases,
while the number of integrand bases is about one hundred for
renormalizable field theories\cite{Feng:2012bm}. Thus the
determination of integral bases for two-loop and higher-loop becomes
necessary.

Although integrand bases can be determined systematically, the
determination of integral bases is far from being completely solved.
It is our attempt in this paper to find an efficient method to solve
the problem. Noticing that in the unitarity method, the action
$\Delta$ in \eref{1loop-exp-1} is directly acting on the integrated
results, thus if the left hand side $\Delta{\cal A}$ can be
analytically integrated for arbitrary inputs,  we can classify
independent distinguishable analytic structures from these results.
Each structure should correspond to one integral basis\footnote{It
is possible that two different integral bases have the same analytic
structure for all physical unitarity cuts, but we do not consider
this possibility in current paper. All our claims in this paper are
true after neglecting  above ambiguity.}. By this attempt we can
determine integral bases.

In this paper, taking double-box topology and its daughter
topologies as examples, we generalize unitarity method to two-loop
amplitudes and try to determine the integral bases. Different from
the maximal unitarity method\cite{Kosower:2011ty}, we cut only four
propagators(the propagator with mixed loop momenta will not be
touched). Comparing with maximal unitarity cut where solutions for
loop momenta are complex number in general, our cut conditions
guarantee the existence of real solutions for loop momenta, thus
avoiding the affects from spurious integrations.

This paper is organized as follows. In section 2 we review the
one-loop unitarity method and then generalize the scheme to
two-loop. For two-loop, two sub-one-loop phase space integrations
should be evaluated. In section 3, we integrate the first
sub-one-loop integration of triangle topology. The result is used in
section 4, where integration over the second sub-one-loop of
triangle topology is performed. Results obtained in this section
allow us to determine integral bases for the topology ${\cal
A}_{212}$. Results in section 3 is also also used in section 5,
where integration over the second sub-one-loop of box topology is
performed, and the result can be used to determine integral bases
for topology ${\cal A}_{213}$. In section 6, we briefly discuss the
integral bases of topology ${\cal A}_{313}$ since results are well
known for this topology. Finally, in section 7, a short conclusion
is given.

Technical details of calculation are presented in Appendix. In
Appendix A, some useful formulae for phase space integration are
summarized. In Appendix B, the phase space integration is done for
one-loop bubble, one-loop triangle and one-loop box topologies. In
Appendix C, details of an integration for topology ${\cal A}_{313}$
are discussed.

\section{Setup}

In this section, we present some general discussions about the
calculation done in this paper. Firstly, we review how to do the
phase space integration in unitarity method illustrated by one-loop
example. Then we set up the framework in unitarity method for
two-loop topologies which are the starting point of this paper.

\subsection{Phase space integration}

The unitarity method has been successfully applied to one-loop
amplitudes
\cite{Bern:1994zx,Bern:1995db,Bern:1997sc,Bern:1996ja,Britto:2004nc,Britto:2004nj,
Bern:2005cq,Britto:2005ha,Brandhuber:2005jw,Bern:2007dw,Forde:2007mi,
Badger:2008cm,Anastasiou:2006jv,Britto:2006fc,Berger:2009zb,Bern:2010qa}
. Here we give a brief summary about the general
$(4-2\epsilon)$-dimensional unitarity method\cite{Anastasiou:2006jv,
Britto:2006fc}, which will be used later. Through this paper we use
the metric $\eta_{\mu\nu}=(+,-,...,-)$ and QCD convention for
spinors, i.e., $2k_i\cdot k_j\equiv \Spaa{k_i|k_j}\Spbb{k_j|k_i}$.

For one-loop, the action $\Delta$ in \eref{1loop-exp-1} is realized
by putting two internal propagators on-shell. More explicitly, let
us consider the following most general input\footnote{The most
general expression for numerator will be $\sum_i \prod_{j}
(\ell\cdot R_{ij})$. For each term $\prod_{j=1}^n (\ell\cdot
R_{ij})$, we can construct $(\ell\cdot \W R_i)^n$ with $\W
R_i=\sum_{j=1}^n y_j R_{ij}$. Thus if we know the result for
numerator $(\ell\cdot \W R_i)^n$, we can expand it into the
polynomial of $y_i$ and read out corresponding result for
$\prod_{j=1}^n (\ell\cdot R_{ij})$. }
 with massless internal propagators\footnote{For simplicity we consider the massless
 propagators, but massive propagators can be dealt similarly. }
\bea {\cal A}^{(a)}_n &\equiv & \int d^{4-2\eps} \WH \ell {\cal
I}_{n }^{(a)} = \int d^{4-2\eps} \WH \ell { (2\WH\ell\cdot T)^a\over
\WH\ell^2\prod_{i=1}^{n-1} (\WH
\ell-K_i)^2}~,~~~\label{1loop-gen-integrand} \eea
where the inner momentum is in $(4-2\epsilon)$-dimensional space and
all external momenta are in pure 4D space for our regularization
scheme. The unitarity cut with intermediate flowing momentum $K$ is
given by putting $\WH \ell^2$ and $(\WH\ell-K)^2$ on-shell, and we
get the expression
\bea \Delta {\cal A}^{(a)}_n  = \int d^{4-2\eps} \WH \ell {
(2\WH\ell\cdot T)^a\delta( \WH\ell^2)\delta((\WH \ell-K)^2)\over
\prod_{i=1}^{n-2} (\WH
\ell-K_i)^2}~.~~~\label{1loop-gen-integrand-cut} \eea
With two delta-functions, the original $(4-2\eps)$-dimensional
integration is reduced to $(2-2\eps)$-dimensional integration. To
carry out the remaining integration, we decompose $\WH\ell$ as
$\WH\ell=\W \ell+\mu$, where $\W\ell$ is the pure 4D part while
$\mu$ is the $(-2\eps)$-dimensional part\cite{Anastasiou:2006jv},
then the measure becomes
\bea \int d^{4-2\eps} \WH \ell\delta( \WH\ell^2)\delta((\WH
\ell-K)^2) (\bullet)=\int d^{-2\epsilon}\mu\int d^4\W \ell\delta(
\W\ell^2-\mu^2) \delta((\W \ell-K)^2-\mu^2)(\bullet)~.~~~ \eea
Next, we split $\W \ell$ into $\W\ell=\ell+zK$ with $\ell^2=0$ to
arrive
\bea & &\int d^4 \W \ell\delta( \W\ell^2-\mu^2)\delta((\W
\ell-K)^2-\mu^2)(\bullet)\nn &=&\int dz
d^4\ell\delta(\ell^2)(2\ell\cdot K)\delta( z^2K^2+2z\ell\cdot
K-\mu^2) \delta((1-2z)K^2-2\ell\cdot
K)(\bullet)~.~~~\label{lightcone} \eea
Having the form \eref{lightcone}, we can use the following well
known result of spinor integration\footnote{For one-loop, we can
take either positive light cone or negative light cone, where for
negative light cone, the $t$-integration will be $\int_{-\infty}^0$.
For two-loop, it can happen that if we take positive light cone for
$\ell_1$, then we need to take negative light cone for $\ell_2$.
However, the choice of light cone only gives an overall sign and
does not affect $\la,\W\la$ integration.}\cite{Cachazo:2004kj}.
Define null momentum as $\ell=t \la \W \la$, then
\bea \int d^4\ell\delta^+(\ell^2)(\bullet)=\int_0^{+\infty}tdt
\int\Spaa{\la|d\la}\Spbb{\W\la|d\W\la}(\bullet)~.~~~\label{mea} \eea
Substituting \eref{mea} back to \eref{lightcone}, we can use
remaining two delta-functions to fix $t$ and $z$ as
%
\bea z={1-\sqrt{1-u}\over 2}~~,~~t={(1-2z)K^2\over
\Spab{\la|K|\W\la}}~~,~~u\equiv {4\mu^2\over K^2}~.~~~\eea

After above simplification, the integral
\eref{1loop-gen-integrand-cut} is transformed to the following
spinor form
\bea \Delta {\cal A}^{(a)}_{n}&=  & \int d^{-2\eps} \mu \int
\Spaa{\la|d\la}\Spbb{\W\la|d\W\la} {(-)^{n-2}[(1-2z) K^2]^{a-n+3}
\Spab{\la|R|\W\la}^a \over
\Spab{\la|K|\W\la}^{a-n+4}\prod_{i=1}^{n-2}\Spab{\la|Q_i|\W\la}}~,~~~
\label{1loop-uni-gen} \eea
where
\bea R \equiv T+ {z (2K\cdot T)\over (1-2z) K^2} K~~,~~Q_i\equiv
K_i+{z (2K\cdot K_i)-K_i^2\over (1-2z) K^2} K~.~~~\label{1loo-R-Q}
\eea
To deal with the integral like $\int
\Spaa{\la|d\la}\Spbb{\W\la|d\W\la}f(\la,\W\la)$ when $f(\la,\W\la)$
is a rational function,
the first step is to find a function $g(\la,\W\la)$ satisfying
\bea \int \Spaa{\la|d\la}\Spbb{\W\la|d\W\la}f(\la,\W\la)= \int
\Spaa{\la|d\la}\Spbb{d\W\la|{\partial \over
\partial\W\la}}g(\la,\W\la)~.~~~ \eea
With $g(\la,\W\la)$, the integration is given algebraically by the
sum of residues of holomorphic pole in
$g(\la,\W\la)$\cite{Britto:2005ha}. In Appendix B, we summarize some
general results of standard one-loop integrations using above
technique. It is worth to mention that for two-loop, $f(\la,\W\la)$
might not be rational function. We will discuss how to deal with it
later.

We also want to remark that under the framework of
$(4-2\eps)$-dimensional unitarity method, coefficient of each basis
will be polynomial of $\mu^2$(remembering  the splitting
$\WH\ell=\W\ell+\mu$). There are two ways to handle it. For the
first way, one can further integrate $\int d^{-2\eps}\mu
~(\mu^2)^{n}$ to find coefficients depending on $\eps$. For the
second way, we just keep $\mu^2$, but include the dimensional
shifted scalar basis\cite{Bern:1996ja,Heinrich:2010ax}, such as
\bea {\cal A}^{D=(4-2\epsilon)}[(\mu^2)^r]\equiv\int
d^{-2\epsilon}\mu d^4 \W\ell{(\mu^2)^r\over
(\W\ell^2-\mu^2)\prod_{i=1}^{n-1}((\ell-K_i)^2-\mu^2)}~.~~~\eea
This is equivalent to
\bea {\cal
A}^{D=(4-2\epsilon)}[(\mu^2)^r]=-\epsilon(1-\epsilon)...(r-1-\epsilon)
{\cal A}^{D=(4+2r-2\epsilon)}[1]~.~~~\eea
For one-loop, dimensional shifted bases are often  used. In this
paper we adapt the similar strategy, i.e., keeping the $\mu$-part
and introducing the dimensional shifted bases.

\subsection{Generalizing to two-loop case}

In this subsection, we set up unitarity method for two-loop
amplitudes, particularly for the attempt of determining integral
bases.

The first problem is to decide which propagators should be cut.
There are three kinds of propagators: (1) propagators depending on
$\WH \ell_1$ only; (2) propagators depending on $\WH \ell_2$ only;
(3) propagators depending on both $\WH \ell_1$ and $\WH \ell_2$. In
principle, we can cut any propagators, but for simplicity, in this
paper we will cut propagators of the first two kinds. For our
choice, we cut two propagators of the first kind and two propagators
of the second kind. With this arrangement, for each loop it is
exactly the familiar unitarity method in one-loop case.

Next we set up notation for two-loop integral. The two internal
momenta are denoted as $\WH \ell_1, \WH \ell_2$ in
$(4-2\eps)$-dimension, while  all external momenta are in pure
4-dimension.  We use $n_1, n_2, n_{12}$ to denote the number of each
kind of propagators respectively. Then a general integrand with
massless propagators\footnote{In this paper, we consider the
massless case only. For inner propagators with masses, we will leave
to further projects.} can  be represented  by\footnote{In this
paper, we use ${\cal I}$ for integrand and ${\cal A}$ for integral.}
\bea {\cal I}_{n_1 n_{12} n_2}^{(a,b)}\equiv { (2\WH\ell_1\cdot
T_1)^a(2\WH\ell_2\cdot T_2)^b\over [\WH\ell_1^2\prod_{i=1}^{n_1-1}
(\WH\ell_1-K_{1i})^2][\WH\ell_2^2\prod_{j=1}^{n_2-1}
(\WH\ell_2-K_{2j})^2][(\WH\ell_1+\WH\ell_2)^2\prod_{t=1}^{n_{12}-1}
(\WH\ell_1+\WH\ell_2-K_t)^2]}~.~~~\label{gen-integrand} \eea
The unitarity cut action $\Delta$ is then given by\footnote{We have
neglected some overall factors in the definition of integration
since it does not matter for our discussion.}
\bea \Delta {\cal A} & = &  \int \prod_{i=1}^2 d^{4-2\eps}\WH \ell_i
\left\{{\cal I}_{n_1 n_{12} n_2}^{(a,b)} \prod_{i=1}^2\WH
\ell_i^2(\WH \ell_i-K_{L_i})^2 \right\} \prod_{i=1}^2\delta(\WH
\ell_i^2)\delta((\WH \ell_i-K_{L_i})^2)~.~~~\label{double-uni-cut}
\eea
%

\EPSFIGURE[ht]{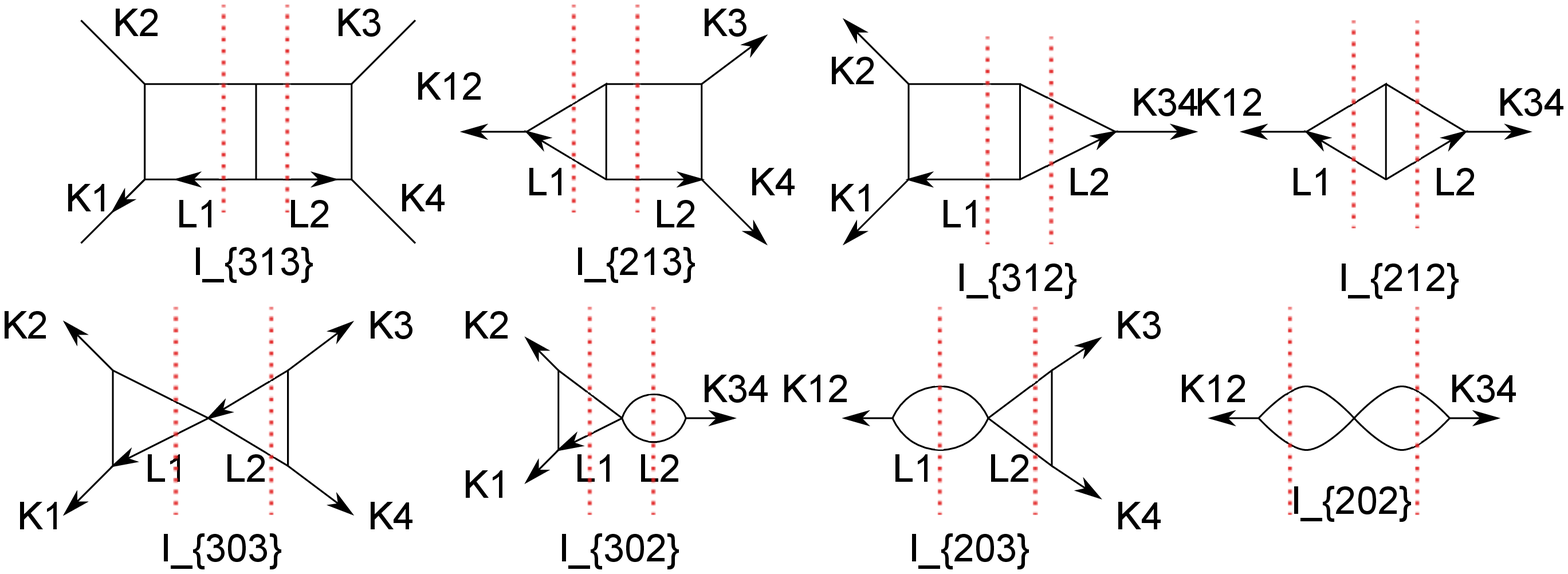,width=17cm}{The unitarity cut of double
box topology $I_{313}$ as well as its seven daughter topologies. The
dashed red lines indicate  cuts. \label{DBox-cut}}

With above setup, we take a well studied example\cite{Gluza:2010ws,
Kosower:2011ty,Larsen:2012sx,CaronHuot:2012ab,
Johansson:2012zv,Johansson:2012sf,Sogaard:2013yga,Johansson:2013sda},
i.e., the four-point two-loop double-box(${\cal A}_{313}$) integral
as the target to apply the unitarity method and determine integral
bases. The integrand is given by
\bea {\cal I}^{(a,b)}_{313} & = & {(2\WH\ell_1 \cdot T_1)^a
(2\WH\ell_2\cdot T_2)^b \over \WH\ell_1^2
(\WH\ell_1-K_1)^2(\WH\ell_1-K_{12})^2 \WH\ell_2^2
(\WH\ell_2-K_4)^2(\WH\ell_2-K_{34})^2
(\WH\ell_1+\WH\ell_2)^2}~,~~~\label{I313-def}\eea
and the four propagators to be cut are
\bean \WH\ell_1^2~~,~~ (\WH\ell_1-K_{12})^2~~,~~ \WH\ell_2^2~~,~~
(\WH\ell_2-K_{34})^2~,~~~\eean
where $K_{12}+K_{34}=0$. With this choice of cuts, in order to
completely understand the results, we also need to consider other
topologies besides double-box. The other contributions come from
those topologies by pinching one or more un-cut propagators of
double-box, as shown in Figure \ref{DBox-cut}. There are three
daughter topologies ${\cal I}_{213},{\cal I}_{312},{\cal I}_{303}$
by pinching one propagator. There are also three daughter topologies
${\cal I}_{212},{\cal I}_{302},{\cal I}_{203}$ by pinching two
propagators. Finally there is only one daughter topology ${\cal
I}_{202}$ by pinching three propagators. Among them, ${\cal
I}_{303}, {\cal I}_{203}, {\cal I}_{302}, {\cal I}_{202}$ are direct
products of two one-loop topologies, thus their signatures are well
known(see Appendix \ref{B}). So in fact we need to examine two
non-trivial topologies ${\cal I}_{212},{\cal I}_{213}$(by symmetry
${\cal I}_{312}$ is equivalent to ${\cal I}_{213}$) together with
the mother topology ${\cal I}_{313}$. Integrand of these two
additional topologies are given by
\bea {\cal I}^{(a,b)}_{212} & = & {(2\WH\ell_1 \cdot T_1)^a
(2\WH\ell_2\cdot T_2)^b \over \WH\ell_1^2 (\WH\ell_1-K_{12})^2
\WH\ell_2^2 (\WH\ell_2-K_{34})^2 (\WH\ell_1+\WH\ell_2)^2}~,~~~\nn
{\cal I}^{(a,b)}_{213} & = &{(2\WH\ell_1 \cdot T_1)^a
(2\WH\ell_2\cdot T_2)^b \over \WH\ell_1^2 (\WH\ell_1-K_{12})^2
\WH\ell_2^2 (\WH\ell_2-K_4)^2(\WH\ell_2-K_{34})^2
(\WH\ell_1+\WH\ell_2)^2}~.~~~\label{I312I213I303}\eea
In the following sections, we will study ${\cal I}_{212}$, ${\cal
I}_{213}$ and ${\cal I}_{313}$ one by one, and our basic strategy
will be to integrate one loop momentum $\W\ell_1$ first while
keeping $\W \ell_2$ arbitrarily. Then we analyze the integration of
$\W\ell_2$ based on the previous results.

\section{The $\W \ell_1$-part integration ($n_1=2$)}

In this section, we do the $\W\ell_1$ integration. Using the
standard method for one-loop amplitudes(reviewed in previous section
as well as in Appendix \ref{B}) we get(see formula
\eref{gen-integrand})
\bea \Delta{\cal A}_{n_1 1 n_2}^{(a,b)} & = & \int d^{-2\eps} \mu_1
d^{-2\eps} \mu_2 d^4 \W\ell_2 \delta(\W\ell_2^2-\mu_2^2)
 \delta(  K_{L_2}^2-2K_{L_2}\cdot \W\ell_2){ (2\W\ell_2\cdot T_2)^b
 \over \prod_{j=1}^{n_2-2} ((\W\ell_2-K_{2j})^2-\mu_2^2)}\nn
 & &  \int  \Spaa{\la_1|d\la_1}\Spbb{\W\la_1|d\W\la_1}
{ (-)^{n_1-2} ((1-2z_1) K_{L_1}^2)^{a-n_1+2}\over
\Spab{\la_1|K_{L_1}|\W\la_1}^{a-n_1+3}}{
\Spab{\la_1|R_1|\W\la_1}^a\over \Spab{\la_1|W_1|\W\la_1}
\prod_{i=1}^{n_1-2}\Spab{\la_1|Q_{1i}|\W\la_1}}~,~~~\label{L1-delta}\eea
where various quantities are defined as
\bea R_1 & \equiv & T_1+{z_1 2K_{L_1}\cdot T_1\over
(1-2z_1)K_{L_1}^2} K_{L_1}~,~~~\nn
Q_{1i} & \equiv & K_{1i}+ {z_1(2 K_{L_1}\cdot K_{1i})-K_{1i}^2\over
(1-2z_1)K_{L_1}^2} K_{L_1}~,~~~\nn
W_1 & \equiv & \W\ell_2+ {(\W\ell_2^2-\mu_2^2)-2\mu_1\cdot
\mu_2+2z_1 \W\ell_2\cdot K_{L_1}\over (1-2z_1)K_{L_1}^2}
K_{L_1}~,~~~\label{L1-var}\eea
with $z_1={1-\sqrt{1-u_1}\over 2}$ and $u_1={4\mu_1^2\over
K_{L_1}^2}$. Note that here the left cut momentum $K_{L_1}=K_{12}$
is the same to the right cut momentum $K_{L_2}=K_{34}$ up to a sign,
however we keep them independently so that it is possible to
formulate them to more general situations. The $W_1$ comes from the
mixed propagator $(\WH \ell_1+\WH \ell_2)^2$. Situations with non
trivial topologies ${\cal A}_{3 1 3}$, ${\cal A}_{3 1 2}$, ${\cal
A}_{2 1 3}$ and ${\cal A}_{2 1 2}$ are all included in the formula
\eref{L1-delta}.

Let us apply our general framework to the specific case $n_1=2$. The
general formula \eref{L1-delta} now becomes
\bea
\Delta{\cal A}_{n_1 1 n_2}^{(a,b)}\Big|_{n_1=2} & = & \int d^{-2\eps} \mu_1
d^{-2\eps} \mu_2 d^4 \W\ell_2 \delta(\W\ell_2^2-\mu_2^2)
 \delta(  K_{L_2}^2-2K_{L_2}\cdot \W\ell_2){ (2\W\ell_2\cdot T_2)^b
 \over \prod_{j=1}^{n_2-2} ((\W\ell_2-K_{2j})^2-\mu_2^2)}\nn
 & &  \int  \Spaa{\la_1|d\la_1}\Spbb{\W\la_1|d\W\la_1}
{  ((1-2z_1) K_{L_1}^2)^{a}\over
\Spab{\la_1|K_{L_1}|\W\la_1}^{a+1}}{ \Spab{\la_1|R_1|\W\la_1}^a\over
\Spab{\la_1|W_1|\W\la_1}}~.~~~\label{L1-delta-n1=2-0}\eea
The second line is nothing but the standard one-loop triangle
integration(see Appendix \ref{B}). When $a=0$, the integration gives
the signature of triangle part. When $a\geq 1$, the integration can
be decomposed into both triangle part and bubble part. We will
evaluate contributions from these two parts separately.

\subsection{The contribution to triangle part}

{\bf The triangle signature :} Based on our general formula of the
standard one-loop triangle integration \eref{C3-to3}, the signature
of the triangle part is
\bea {\cal S}_{tri}\equiv {1\over
\sqrt{\Delta_{W_1,K_{L_1}}}}\ln\left({ W_1 \cdot K_{L_1}- \sqrt{
(W_1\cdot K_{L_1})^2 -W_1^2 K_{L_1}^2}\over  W_1 \cdot K_{L_1}+
\sqrt{  (W_1\cdot K_{L_1})^2 -W_1^2 K_{L_1}^2}}\right)~.~~~ \eea
Imposing cut conditions for $\W \ell_2$, i.e.,
$\delta(\W\ell_2^2-\mu_2^2)$ and $
 \delta(  K_{L_2}^2-2K_{L_2}\cdot \W\ell_2)$ we can simplify it to
\bea {\cal S}_{tri}&= &{1\over K_{L_1}^2 \sqrt{1-u_2}} \ln \left( {
(4\mu_1\cdot \mu_2+K_{L_1}^2)+\sqrt{(1-u_1)(1-u_2)} K_{L_1}^2\over
(4\mu_1\cdot \mu_2+K_{L_1}^2)-\sqrt{(1-u_1)(1-u_2)}
K_{L_1}^2}\right)={1\over t_2K_{L_1}^2 } \ln \Big( {s+t_1t_2\over
s-t_1t_2}\Big)~,~~~\label{3to3-sign} \eea
where we have introduced
\bea s= {4\mu_1\cdot \mu_2+K_{L_1}^2\over
K_{L_1}^2}~~,~~t_i=\sqrt{1-u_i}~~,~~u_i={4\mu_i^2\over
K_{L_i}^2}~~,~~i=1,2~.~~~ \label{combi}\eea
One can observe that the signature part does not depend on
$\W\ell_2$. It is an important feature which makes $\Delta{\cal
A}_{2 1 n_2}^{(a,b)}$ easier to be treated.

~\\{\bf The coefficient ${\cal C}_{3\to 3}^{(a)}$:} Using
\eref{C3-to3} the expression is
\bea {\cal C}_{3\to 3}^{(a)}&=&{(-)^a\over
a!\Delta_{W_1,K_{L_1}}^a}{d^a\over d\tau^a}\Big(\tau^2W_1^2+\tau(4
W_1^2(R_1\cdot K_{L_1})-4 (R_1\cdot W_1) (W_1\cdot K_{L_1}))+ R_1^2
\Delta_{W_1,K_{L_1}}\nn & &+ (2 R_1\cdot W_1)^2 K_{L_1}^2 +
(2R_1\cdot K_{L_1})^2 W_1^2 - (2R_1\cdot W_1)(2R_1\cdot K_{L_1})
(2W_1\cdot K_{L_1})\Big)^a\Big|_{\tau\to 0}~.~~~\eea
Again, using cut conditions $\delta(\W\ell_2^2-\mu_2^2)$ and $
 \delta(  K_{L_2}^2-2K_{L_2}\cdot \W\ell_2)$ we can do the following replacement
\bean \W\ell_2\to {(1-2z_2) K_{L_2}^2\over
\Spab{\la_2|K_{L_2}|\W\la_2}}\la_2 \W\la_2+z_2 K_{L_2}={(1-2z_2)
K_{L_1}^2\over -\Spab{\la_2|K_{L_1}|\W\la_2}}\la_2 \W\la_2-z_2
K_{L_1}~,~~~\eean
where $z_2={1-t_2\over 2}$. Since all derivatives act on $\tau$
only, such replacement will not affect the result. Some algebraic
manipulations shows that the coefficients of different parts are
given by
\bea \tau^2&:&~~{ (s^2-t_1^2t_2^2)K_{L_1}^2\over 4 t_1^2}~,~~~\nn
\tau&:&  {-t_2K_{L_1}^2 \over t_1\Spab{\la_2|K_{L_1}|\W\la_2}}\Big(
t_2
 (K_{L_1}\cdot T_1)
\Spab{\la_2|K_{L_1}|\W\la_2}+s( -(K_{L_1}\cdot
T_1)\Spab{\la_2|K_{L_1}|\W\la_2}+K_{L_1}^2
\Spab{\la_2|T_1|\W\la_2})\Big)~,~~~\nn \tau^0 &: & {t_2^2
(K_{L_1}^2)^2\over \Spab{\la_2|K_{L_1}|\W\la_2}^2}
K_{L_1}^2\Spab{\la_2|(T_1+y_1 K_{L_1})|\W\la_2}\Spab{\la_2|(T_1+y_2
K_{L_1})|\W\la_2}~,~~~\label{n1=2-tau} \eea
with
\bea y_{1,2}={ -(2T_1\cdot K_{L_1})\pm \sqrt{(2T_1\cdot K_{L_1})^2-4
K_{L_1}^2 T_1^2}\over 2K_{L_1}^2}~.~~~\label{y1y2} \eea
To get non-zero contribution from ${d^a\over d\tau^a}
(\bullet)\Big|_{\tau\to 0}$, we only need to take terms with
$\tau^a$ power. It means that terms with $\tau^2$ in \eref{n1=2-tau}
will always appear with terms $\tau^0$, therefore we can regroup
\bean \left\{\tau^2{ (s^2-t_1^2t_2^2)K_{L_1}^2\over 4
t_1^2}\right\}+\left\{{t_2^2 (K_{L_1}^2)^2\over
\Spab{\la_2|K_{L_1}|\W\la_2}^2} K_{L_1}^2\Spab{\la_2|(T_1+y_1
K_{L_1})|\W\la_2}\Spab{\la_2|(T_1+y_2 K_{L_1})|\W\la_2}\right\}
\eean
to
\bean \left\{\tau^2{ t_2(s-t_1t_2)(K_{L_1}^2)^2\over 2
t_1}{\Spab{\la_2|(T_1+y_1 K_{L_1})|\W\la_2}\over
\Spab{\la_2|K_{L_1}|\W\la_2}} \right\}+\left\{{
t_2(s+t_1t_2)(K_{L_1}^2)^2\over 2 t_1}{\Spab{\la_2|(T_1+y_2
K_{L_1})|\W\la_2}\over \Spab{\la_2|K_{L_1}|\W\la_2}}\right\}~.~~~
\eean
Thus we can write
\bean {\cal C}_{3\to 3}^{(a)} &=&{(-)^a (K_{L_1}^2)^a\over a!
(t_1t_2K_{L_1}^2)^a} {d^a\over d\tau^a} {\Spab{\la_2|{\cal
F}|\W\la_2}^a\over \Spab{\la_2|K_{L_1}|\W\la_2}^a}\Big |_{\tau\to
0}~,~~~\eean
where ${\cal F}$ is defined as
\bea {\cal F} & = & -\tau\left( t_2
 {K_{L_1}\cdot T_1\over K_{L_1}^2} K_{L_1}+ s\Big(T_1- {(K_{L_1}\cdot T_1)\over K_{L_1}^2}
K_{L_1}\Big)\right)\nn & &  +\tau^2 { s-t_1t_2\over 2 }\Big(T_1+y_1
K_{L_1}\Big)+ { s+t_1t_2\over 2 }\Big(T_1+y_2
K_{L_1}\Big)~.~~~\label{n1=2-F-def} \eea
Putting all results together, the triangle part becomes
\bea {\cal R}_{3\to 3}^{(a)} & = & \left({(-)^a(K_{L_1}^2)^a\over a!
(t_1t_2K_{L_1}^2)^a} {d^a\over d\tau^a} {\Spab{\la_2|{\cal
F}|\W\la_2}^a\over \Spab{\la_2|K_{L_1}|\W\la_2}^a}\Big|_{\tau\to
0}\right){1\over t_2K_{L_1}^2 } \ln \Big( {s+t_1t_2\over
s-t_1t_2}\Big)~.~~~\label{n1=2-R3-3} \eea
To do the $\W \ell_2$-part integration, it is more convenient to use
above form  before taking the derivative over $\tau$.

\subsection{The contribution to bubble part}
Again we use results given in Appendix \ref{B}.

{\bf The ${\cal R}_{3\to 2}[i,m]$ term:} Using \eref{R-3-to-2}, the
typical term of triangle topology to bubble is
\bea
{\cal R}_{3\to 2}[i,m] & = & { (-)^{m+i} (K_{L_1}^2)^i \over
i!(m+1)\sqrt{\Delta(W_1,K_{L_1})}^{m+2i+2}}{d^i\over d\tau^i}\left\{\left((2
R_1\cdot P_2 -\tau\Spab{P_1|R_1|P_2})^{m+1}\right.
 \right.\nn & & \left.
(-x_2 \Spab{P_2|R_1|P_1}-x_1 \tau^2 \Spab{P_1|R_1|P_2}+\tau ( x_2
(2R_1\cdot P_1) +x_1(2R_1\cdot P_2)))^i\right)\nn & & +(-)^m
\left((2 R_1\cdot P_1-\tau\Spab{P_2|R_1|P_1})^{m+1}\right. \nn & &
\left.\left. (-x_2\tau^2\Spab{P_2|R_1|P_1} -x_1 \Spab{P_1|R_1|P_2}
+\tau ( x_2 (2R_1\cdot P_1) +x_1(2R_1\cdot
P_2)))^i\right)\right\}\Big|_{\tau\to 0}~,~~~\label{R1-3-to-2} \eea
where two null momenta $P_1,~P_2$ are constructed as
$P_i=W_1+x_iK_{L_1}$, with
\bean x_1={s+t_1t_2\over 2t_1}~~,~~x_2={s-t_1t_2\over
2t_2}~.~~~\eean
Again, to get non-zero contribution, $\Spab{P_1|R_1|P_2}$ and
$\Spab{P_2|R_1|P_1}$ should always appear in pair. With a little
calculations, one can see
\bea \Spab{P_1|R_1|P_2}\Spab{P_2|R_1|P_1}={\cal T}_1{\cal T}_2~~,~~
{\cal T}_i= \left({t_2K_{L_1}^2\over
\Spab{\la_2|K_{L_1}|\W\la_2}}\right)\Spab{\la_2|(T_1+y_i
K_{L_1})|\W\la_2}~,~~~\label{n1=2=rep-T}\eea
where $y_1$ and $y_2$ are defined in \eref{y1y2}. Thus we can take
the following replacements
\bea \Spab{P_1|R_1|P_2}&\to & {\cal T}_1~~,~~
\Spab{P_2|R_1|P_1}\to {\cal T}_2~.~~~\eea
After such replacements we obtain
\bea
& & {\cal R}_{3\to 2}[i,m]  =  { (-)^{m+i}  \over
(m+1)i!K_{L_1}^2 t_2^{i+1} t_1^{m+i+1}}{d^i\over
d\tau^i} \nn
& &\left\{ { \Spab{\la_2|T_1(-t_1-\tau t_1)+K_{L_1}(-\tau t_1 y_1-(1-t_1){K_{L_1}\cdot T_1\over
K_{L_1}^2}) |\W\la_2}^{m+1}\over \Spab{\la_2|K_{L_1}|\W\la_2}^{m+i+1}} \right. \nn
& & \times \Spab{\la_2|T_1 (-x_2 t_1-x_1 \tau^2 t_1-s\tau)+K_{L_1}(
-x_2 t_1 y_2-x_1 t_1 \tau^2 y_1+\tau (s-t_2){K_{L_1}\cdot T_1\over
K_{L_1}^2})  | \W\la_2}^{i}\nn
& & +(-)^m { \Spab{\la_2|T_1(-t_1-\tau t_1)+K_{L_1}(-\tau t_1 y_2+(1+t_1){K_{L_1}\cdot T_1\over
K_{L_1}^2}) |\W\la_2}^{m+1}\over \Spab{\la_2|K_{L_1}|\W\la_2}^{m+i+1}}  \nn
& & \left.\times \Spab{\la_2|T_1 (-x_2 t_1\tau^2-x_1
t_1-s\tau)+K_{L_1}( -x_2 t_1\tau^2 y_2-x_1 t_1  y_1+\tau
(s-t_2){K_{L_1}\cdot T_1\over K_{L_1}^2})  |
\W\la_2}^{i}\right\}\Big|_{\tau\to 0}~.~~~ \eea
Above expression has the form
$\Spab{\la_2|\bullet|\W\la_2}\Spab{\la_2|\bullet|\W\la_2}$. In order
to use the results given in Appendix \ref{B}, we need to rewrite
them by using
\bean
 (A)^{m+1}(B)^i= {i!\over (m+1+i)!}{d^{m+1}\over d \tau_1^{m+1}}(\tau_1 A+ B)^{m+1+i}\Big|_{\tau_1\to
 0}~.~~~
\eean
So finally we have
\bea {\cal R}_{3\to 2}[i,m]  &=&  { (-)^{m+i}  \over
(m+1)(m+1+i)!K_{L_1}^2 t_2^{i+1}t_1^{m+i+1}}\left\{{d^i\over
d\tau^i}{d^{m+1}\over d\tau_1^{m+1}}\right.\nn & & \left.
{\Spab{\la_2|{\cal G}_2|\W\la_2}^{m+i+1}+(-)^m \Spab{\la_2|{\cal
G}_1|\W\la_2}^{m+i+1}\over
\Spab{\la_2|K_{L_1}|\W\la_2}^{m+i+1}}\right\}_{\tau\to 0,\tau_1\to
0}~,~~~\eea
where we have defined
\bea {\cal G}_1 & = & T_1 \left\{ -t_1\tau_1-{(s-t_1 t_2)\over 2}
\tau^2-{(s+t_1 t_2)\over 2}-\tau s-\tau \tau_1 t_1\right\}\nn
& & +K_{L_1}\left\{\tau_1 {K_{L_1}\cdot T_1\over K_{L_1}^2} (1+t_1)-
\tau^2  y_2{(s-t_1 t_2)\over 2}-{(s+t_1 t_2)\over 2} y_1+ \tau
{K_{L_1}\cdot T_1\over K_{L_1}^2}(s-t_2)-\tau \tau_1 t_1
y_2\right\}~,~~~ \nn {\cal G}_2 & = & T_1\left\{ -\tau_1 t_1-\tau
\tau_1 t_1-{(s-t_1 t_2)\over 2}-\tau^2 {(s+t_1 t_2)\over
2}-s\tau\right\}\nn
& & +K_{L_1}\left\{ [-(1-t_1) \tau_1+\tau (s-t_2)]{K_{L_1}\cdot
T_1\over K_{L_1}^2}-\tau \tau_1 t_1 y_1 -{(s-t_1 t_2)\over
2}y_2-\tau^2 {(s+t_1 t_2)\over 2}y_1\right\}~.~~~\label{cal-G1-G2}
\eea
%

\subsection{The  result for $n_1=2$ after $\W \ell_1$-integration}

Collecting results from triangle part and bubble part we obtain
\bea
\Delta{\cal A}_{n_1 1 n_2}^{(a,b)}\Big|_{n_1=2} & = & \int
d^{-2\eps} \mu_1 d^{-2\eps} \mu_2 \int d^4 \W\ell_2
\delta(\W\ell_2^2-\mu_2^2)
 \delta(  K_{L_2}^2-2K_{L_2}\cdot \W\ell_2){ (2\W\ell_2\cdot T_2)^b(t_1
 K_{L_1}^2)^{a}
 \over \prod_{j=1}^{n_2-2} ((\W\ell_2-K_{2j})^2-\mu_2^2)}\nn
 & & \left\{{1\over K_{L_1}^2 t_2} \ln \left( { (s+ t_1 t_2)\over (s-t_1 t_2)}\right) {(-)^a (K_{L_1}^2)^a\over a! (t_1 t_2 K_{L_1}^2)^a} {d^a\over
d\tau^a} {\Spab{\la_2|{\cal F}|\W\la_2}^a\over
\Spab{\la_2|K_{L_1}|\W\la_2}^a}\Big|_{\tau\to
0}\right.~~~\label{n1=2-la1}\\
& &\left. +\sum_{i=0}^{a-1} { (-)^{a-1}  \over (a-i)a!K_{L_1}^2
t_2^{i+1}t_1^{a}}{d^i\over d\tau^i}{d^{a-i}\over d\tau_1^{a-i}}
{\Spab{\la_2|{\cal G}_2|\W\la_2}^{a}+(-)^{a-1-i} \Spab{\la_2|{\cal
G}_1|\W\la_2}^{a}\over
\Spab{\la_2|K_{L_1}|\W\la_2}^{a}}\Big|_{\tau\to 0,\tau_1\to
0}\right\}~,~~~\nonumber \eea
where $s,t_1,t_2$ are defined in \eref{combi}, ${\cal F}$ in
\eref{n1=2-F-def} and ${\cal G}_1, {\cal G}_2$ in
\eref{cal-G1-G2}\footnote{Do not confuse the $t_2$ here with the
$t_2$-integration part of $\W\ell_2$ as reviewed in \eref{mea}.}.
The trick here is that instead of computing the operations
${d^a\over d\tau^a} (\bullet)\Big|_{\tau\to 0}$ and ${d^i\over
d\tau^i}{d^{a-i}\over d\tau_1^{a-i}} (\bullet)\Big|_{\tau\to
0,\tau_1\to 0}$, we will firstly do the $\W \ell_2$-part
integration.

For $\W\ell_2$-integration, after the $t_2$-integration we are left
with spinor integration given by\footnote{There is an overall sign
for $t_2$-integration since  the momentum conservation forces
$K_{L_1}=-K_{L_2}$, i.e.,   $\Spab{\la_2|K_{L_2}|\W\la_2}<0$.}
\bea & &\Delta{\cal A}_{n_1 1 n_2}^{(a,b)}\Big|_{n_1=2} =  \int
d^{-2\eps} \mu_1 d^{-2\eps} \mu_2 \int \Spaa{\la_2|d\la_2}
\Spbb{\W\la_2|d\W\la_2}{ (-)^{n_2+1} \Spab{\la_2|R_2|\W\la_2}^{b}
 \over \prod_{j=1}^{n_2-2} \Spab{\la_2|Q_{2j}|\W\la_2}
 \Spab{\la_2|K_{L_2}|\W\la_2}^{2+b-(n_2-2)}} \nn & &
   \left\{{(-)^a t_2^{b-(n_2-2)}(K_{L_1}^2)^{a+b-(n_2-2)}\over a! t_2^a} \ln \left( { (s+ t_1 t_2)\over (s-t_1 t_2)}\right) {d^a\over
d\tau^a} {\Spab{\la_2|{\cal F}|\W\la_2}^a\over
\Spab{\la_2|K_{L_1}|\W\la_2}^a}\Big|_{\tau\to
0}\right.~~~\label{n1=2-la2}\\
& &\left. +\sum_{i=0}^{a-1} { (-)^{a-1} (K_{L_1}^2)^{a+b-(n_2-2)}
t_2^{b-(n_2-2)}  \over (a-i)a! t_2^{i}}{d^i\over
d\tau^i}{d^{a-i}\over d\tau_1^{a-i}} {\Spab{\la_2|{\cal
G}_2|\W\la_2}^{a}+(-)^{a-1-i} \Spab{\la_2|{\cal
G}_1|\W\la_2}^{a}\over
\Spab{\la_2|K_{L_1}|\W\la_2}^{a}}\Big|_{\tau\to 0,\tau_1\to
0}\right\}~,~~~\nonumber \eea
where we have defined
\bea R_2  \equiv  T_2+{z_2 2K_{L_2}\cdot T_2\over (1-2z_2)K_{L_2}^2}
K_{L_2}~~,~~ Q_{2j} \equiv  K_{2j}+ {z_2(2 K_{L_2}\cdot
K_{2j})-K_{2j}^2\over (1-2z_2)K_{L_2}^2} K_{L_2}~.~~~\label{L2-var}
\eea
%

\section{The integral bases of ${\cal A}_{212}$ topology}

With results of previous section, it is possible to discuss the
integral bases of ${\cal A}_{212}$ topology in this section. To do
so, we need to finish the spinor integration given in
\eref{n1=2-la2} with $n_2=2$, and attempt to identify the results to
the bases. We will see that there are only (dimensional shifted)
scalar bases.

\subsection{The $\la_2$-integration for the case $n_2=2$}

For the case $n_2=2$ the formula \eref{n1=2-la2} becomes
\bea
\Delta{\cal A}_{2 1 2}^{(a,b)}& = & \int
d^{-2\eps} \mu_1 d^{-2\eps} \mu_2 \int \Spaa{\la_2|d\la_2}
\Spbb{\W\la_2|d\W\la_2}{  (-)\Spab{\la_2|R_2|\W\la_2}^{b}
 \over
 \Spab{\la_2|K_{L_2}|\W\la_2}^{2+b}} \nn & &
   \left\{{(-)^a (K_{L_1}^2)^{a+b}t_2^{b}\over a! t_2^a} \ln \left( { (s+ t_1 t_2)\over (s-t_1 t_2)}\right) {d^a\over
d\tau^a} {\Spab{\la_2|{\cal F}|\W\la_2}^a\over
\Spab{\la_2|K_{L_1}|\W\la_2}^a}\Big|_{\tau\to
0}\right.~~~\label{n1=2-la2-2}\\
& &\left. +\sum_{i=0}^{a-1} { (-)^{a-1} (K_{L_1}^2)^{a+b} t_2^{b}
\over (a-i)a! t_2^{i}}{d^i\over d\tau^i}{d^{a-i}\over d\tau_1^{a-i}}
{\Spab{\la_2|{\cal G}_2|\W\la_2}^{a}+(-)^{a-1-i} \Spab{\la_2|{\cal
G}_1|\W\la_2}^{a}\over
\Spab{\la_2|K_{L_1}|\W\la_2}^{a}}\Big|_{\tau\to 0,\tau_1\to
0}\right\}~.~~~\nonumber \eea
For our momentum configuration, $K_{L_1}=-K_{L_2}$, thus we can
combine denominator together to get a simpler expression. Terms of
integrand can be classified into two parts, and we evaluate them one
by one.

~\\ {\bf First part:} The first part can be rewritten as
\bea
& & \int
d^{-2\eps} \mu_1 d^{-2\eps} \mu_2 {(-)^{a+b+1} (K_{L_1}^2)^{a+b}t_2^{b}\over a! t_2^a} \ln \left( { (s+ t_1 t_2)\over (s-t_1 t_2)}\right) {d^a\over
d\tau^a}{ a!\over (a+b)!}{d^b\over
d\W\tau^b}\nn & &  \int \Spaa{\la_2|d\la_2}
\Spbb{\W\la_2|d\W\la_2}{  \Spab{\la_2|\W\tau R_2+{\cal F}|\W\la_2}^{a+b}
 \over
 \Spab{\la_2|K_{L_2}|\W\la_2}^{a+b+2}}\Big|_{\tau\to
0,\W\tau\to 0}~.~~~ \eea
The second line is the standard one-loop bubble integration, thus we
can use the general formulae in Appendix \ref{B}.

~\\ {\bf Second part:} The second part can be rewritten as
\bea & & \int d^{-2\eps} \mu_1 d^{-2\eps} \mu_2 \sum_{i=0}^{a-1} {
(-)^{a+b} (K_{L_1}^2)^{a+b} t_2^{b}  \over (a-i)(a+b)!
t_2^{i}}{d^b\over d\W\tau^b}{d^i\over d\tau^i}{d^{a-i}\over
d\tau_1^{a-i}}\nn & &  \int \Spaa{\la_2|d\la_2}
\Spbb{\W\la_2|d\W\la_2} {\Spab{\la_2|\W \tau R_2+{\cal
G}_2|\W\la_2}^{a+b}+(-)^{a-1-i} \Spab{\la_2|\W \tau R_2+ {\cal
G}_1|\W\la_2}^{a+b}\over
\Spab{\la_2|K_{L_1}|\W\la_2}^{a+b+2}}\Big|_{\tau\to 0,\tau_1\to
0,\W\tau\to 0}~.~~~ \eea
The second line is again the one-loop bubble integration. After
finishing the integration over $\la_2$-part, we can take the
derivative and the limit $\tau\to 0,\tau_1\to 0,\W\tau\to 0$.

\subsection{The result}

Collecting all results together, we get an expression of the form
\bea & & \Delta{\cal A}_{2 1 2}^{(a,b)}  =  \int d^{-2\eps} \mu_1
d^{-2\eps} \mu_2\left\{ f_{212\to 202}^{(a,b)} {\cal
S}_{202}+f_{212\to 212}^{(a,b)}{\cal S}_{212} \right\}~,~~~
\label{A212-final-0} \eea
where we have defined
\bea {\cal S}_{202}= -t_1t_2~~,~~{\cal S}_{212}={1\over
K_{L_1}^2}\ln{\left({s+t_1t_2\over
s-t_1t_2}\right)}~.~~~\label{212-sgn} \eea
Remind from Appendix \ref{B} that the signature of one-loop bubble
is $\int d^{-2\eps} \mu (-\sqrt{1-u^2})$, thus the term ${\cal
S}_{202}$ is the signature of topology ${\cal A}_{202}$ as the
subscript indicates. For ${\cal S}_{212}$, since the factor
$\ln{\left({s+t_1t_2\over s-t_1t_2}\right)}$ can not be factorized
to a form where $\mu_1$-part and $\mu_2$-part are decoupled,  it can
not belong to the topology ${\cal A}_{n_1 0 n_2}$. So it must be the
signature of topology ${\cal A}_{212}$.

It is worth to mention that in the form \eref{A212-final-0}, the
dependence of $a,b$ is completely encoded in the coefficients
$f_{212\to 202}^{(a,b)}$ and $f_{212\to 212}^{(a,b)}$, while the
signature \eref{212-sgn} is universal.  However, it does not mean
the basis is just given by $a=b=0$. It could be  true  only when
coefficients $f_{212\to 202}^{(a,b)}$ and $f_{212\to 212}^{(a,b)}$
satisfying the following two conditions: (1) they are  polynomials
of $u_1, u_2$ and $s$; (2) they are rational functions of external
momentum $K_{L_1}$. More discussions will be given shortly after.

Having above general discussions, now we list coefficients for
various $a,b$:

~\\ {\bf Coefficients $f_{212\to 212}$:} Using expression given in
Appendix \ref{B}, the analytic results for some levels of $a+b$ are
given by ~\\{\bf $\bullet$ $a+b=0,1$:}
\bea f_{212\to 212}^{(0,0)} & = & 1~~,~~f_{212\to 212}^{(1,0)}= {
T_1\cdot K_{L_1}}~~,~~f_{212\to 212}^{(0,1)}= { -T_2\cdot
K_{L_1}}~.~~~ \eea
~\\ {\bf $\bullet$ $a+b=2$:}
\bea f_{212\to 212}^{(1,1)} & = & {1\over 3} \left( s K_{L_1}^2
(T_1\cdot T_2) -(3+s) (K_{L_1}\cdot T_1) (K_{L_1}\cdot T_2)
\right)~,~~~\nn
f_{212\to 212}^{(2,0)} & = &{1\over 3} \left(  (3+ (1-u_1))
(K_{L_1}\cdot T_1)^2 -(1-u_1) K_{L_1}^2 T_1^2\right)~,~~~ \nn
f_{212\to 212}^{(0,2)}&= & {1\over 3} \left(  (3+ (1-u_2))
(K_{L_1}\cdot T_2)^2 -(1-u_2) K_{L_1}^2 T_2^2\right)~.~~~ \eea
~\\ {\bf $\bullet$ $a+b=3$:}
\bea f_{212\to 212}^{(1,2)} & = & {1\over 3} \left( -2s K_{L_1}^2
(K_{L_1}\cdot T_2) (T_1\cdot T_2) + (K_{L_1}\cdot T_1)( (3+2
s+(1-u_2)) (K_{L_1}\cdot T_2)^2\right.\nn & &\left. -(1-u_2)
K_{L_1}^2 T_2^2)\right)~,~~~\nn
f_{212\to 212}^{(0,3)} & = & -(1+(1-u_2)) (K_{L_1}\cdot
T_2)^3+(1-u_2) K_{L_1}^2 (K_{L_1} \cdot T_2) T_2^2~.~~~ \eea
~\\ {\bf $\bullet$ $a+b=4$:}
\bea & & f_{212\to 212}^{(2,2)}  =  {1\over 15}\left\{ 2(-s(10+3 s)+
(1-u_2)(1-u_1)) K_{L_1}^2 (K_{L_1}\cdot T_1)(K_{L_1}\cdot
T_2)(T_1\cdot T_2)\right. \nn & & + (K_{L_1}\cdot T_1)^2 ( (2
s(10+s)+5(3+(1-u_1))+(5+(1-u_1)) (1-u_2)) (K_{L_1}\cdot T_2)^2\nn &
& + (s^2-(5+ 2 (1-u_1)) (1-u_2)) K_{L_1}^2 T_2^2) + K_{L_1}^2 ((
s^2-(1-u_1)(5+2 (1-u_2))) (K_{L_1}\cdot T_2)^2 T_1^2\nn & &
\left.+K_{L_1}^2 ((3 s^2-(1-u_1)(1-u_2)) (T_1\cdot T_2)^2- (s^2-2
(1-u_1)(1-u_2)) T_1^2 T_2^2)) \right\}~.~~~\label{f212-212} \eea

~\\ {\bf Coefficients $f_{212\to 202}$:} ~\\ {$\bullet$ $a=0$ or
$b=0$:} From our derivation, it can easily be seen that when $a=0$
or $b=0$, the coefficient must be zero, i.e.,
\bea f_{212\to 202}^{(0,b)}=f_{212\to 202}^{(a,0)}=0~.~~~\eea
~\\ {$\bullet$ Non-zero results:}
\bea f_{212\to 202}^{(1,1)} & = & {2\over 3} \left( T_1\cdot T_2-{
(K_{L_1}\cdot T_1)(K_{L_1}\cdot T_2)\over K_{L_1}^2}\right)~,~~~\nn
f_{212\to 202}^{(1,2)} & = & { 4 (K_{L_1}\cdot T_2) ((K_{L_1}\cdot
T_1)(K_{L_1}\cdot T_2)- K_{L_1}^2(T_1\cdot T_2))\over 3
K_{L_1}^2}~,~~~ \nn
f_{212\to 202}^{(1,3)} & = & {2 ((K_{L_1}\cdot T_1)(K_{L_1}\cdot T_2)- K_{L_1}^2(T_1\cdot T_2))
( (5+(1-u_2)) (K_{L_1}\cdot T_2)^2-(1-u_2) K_{L_1}^2 T_2^2)
 \over -5 K_{L_1}^2}~,~~~\nn
f_{212\to 202}^{(2,2)} & = & {2\over 15 K_{L_1}^2} \left\{ -2(10+3
s) K_{L_1}^2 (K_{L_1}\cdot T_1) (K_{L_1}\cdot T_2)(T_1 \cdot T_2)+
(K_{L_1}\cdot T_1)^2( 2(10+s) (K_{L_1}\cdot T_2)^2 \right. \nn & &
\left.+s K_{L_1}^2 T_2^2)+s K_{L_1}^2 ( (K_{L_1}\cdot T_2)^2
T_1^2+K_{L_1}^2 (3 (T_2\cdot T_1)^2- T_1^2 T_2^2)) \right\}~.~~~\eea
%

\subsection{Classification of integral bases}

Now we need to analyze above results in order to determine the
integral bases. Firstly, noticing that $f_{212\to 212}^{(a,b)}$ and
$f_{212\to 202}^{(a,b)}$ are polynomials of $T_1, T_2, \mu_1\cdot
\mu_2, \mu_1^2, \mu_2^2$ as well as rational functions of external
momentum $K_{L_1}$, thus we can write them more explicitly as
\bea f_{212\to 212}^{(a,b)} & = & \sum_{\kappa_0,\kappa_1,\kappa_2}
f_{212\to 212; \mu_1,...,\mu_{a};\nu_1,...,\nu_b}^{(a,b)}
T_1^{\mu_1}...T_1^{\mu_a} T_2^{\nu_1}... T_2^{\nu_b}
(\mu_1^2)^{\kappa_1} (\mu_2^2)^{\kappa_2} (\mu_1\cdot
\mu_2)^{\kappa_0}~,~~~\nn
f_{212\to 202}^{(a,b)} & = & \sum_{\kappa_0,\kappa_1,\kappa_2}
f_{212\to 202; \mu_1,...,\mu_{a};\nu_1,...,\nu_b}^{(a,b)}
T_1^{\mu_1}...T_1^{\mu_a} T_2^{\nu_1}... T_2^{\nu_b}
(\mu_1^2)^{\kappa_1} (\mu_2^2)^{\kappa_2} (\mu_1\cdot
\mu_2)^{\kappa_0}~,~~~\label{f212-coeff-exp} \eea
where the tensor coefficients $f_{212\to 212;
\mu_1,...,\mu_{a};\nu_1,...,\nu_b}^{(a,b)}$ are rational functions
of external momentum $K_{L_1}$ only. Putting it back we get
\bea & & \Delta{\cal A}_{2 1 2}^{(a,b)}\nn & = &
\sum_{\kappa_0,\kappa_1,\kappa_2}f_{212\to 202;
\mu_1,...,\mu_{a};\nu_1,...,\nu_b}^{(a,b)} T_1^{\mu_1}...T_1^{\mu_a}
T_2^{\nu_1}... T_2^{\nu_b} \int d^{-2\eps} \mu_1 d^{-2\eps}
\mu_2(\mu_1^2)^{\kappa_1} (\mu_2^2)^{\kappa_2} (\mu_1\cdot
\mu_2)^{\kappa_0}  {\cal S}_{202}\nn & +&
\sum_{\kappa_0,\kappa_1,\kappa_2}f_{212\to 212;
\mu_1,...,\mu_{a};\nu_1,...,\nu_b}^{(a,b)} T_1^{\mu_1}...T_1^{\mu_a}
T_2^{\nu_1}... T_2^{\nu_b}  \int d^{-2\eps} \mu_1 d^{-2\eps}
\mu_2(\mu_1^2)^{\kappa_1} (\mu_2^2)^{\kappa_2} (\mu_1\cdot
\mu_2)^{\kappa_0} {\cal S}_{212}~.~~~ \label{A212-final-1} \eea
The above expansion leads us to define following {\bf dimensional
shifted bases}
\bea {\cal B}_{202}^{(0,0)}[\kappa_0,\kappa_1,\kappa_2] & \equiv &
\int d^{4-2\eps}\WH\ell_1\int d^{4-2\eps}\WH\ell_2
{(\mu_1^2)^{\kappa_1} (\mu_2^2)^{\kappa_2} (\mu_1\cdot
\mu_2)^{\kappa_0} \over \WH\ell_1^2 (\WH\ell_1-K_{L_1})^2
\WH\ell_2^2 (\WH\ell_2+K_{L_1})^2 }\label{B202-dim-basis} \eea
and
\bea {\cal B}_{212}^{(0,0)}[\kappa_0,\kappa_1,\kappa_2] & \equiv &
\int d^{4-2\eps}\WH\ell_1\int d^{4-2\eps}\WH\ell_2
{(\mu_1^2)^{\kappa_1} (\mu_2^2)^{\kappa_2} (\mu_1\cdot
\mu_2)^{\kappa_0} \over \WH\ell_1^2 (\WH\ell_1-K_{L_1})^2
\WH\ell_2^2 (\WH\ell_2+K_{L_1})^2
(\WH\ell_1+\WH\ell_2)^2}~.~~~\label{B212-dim-basis} \eea
An important observation is that in the definition of ${\cal
B}_{202}^{(0,0)}[\kappa_0,\kappa_1,\kappa_2]$, when $\kappa_0\neq
0$, we do have $\mu_1\cdot \mu_2$ in the numerator. Thus although
there is no mixed propagator in the denominator, it contains
information from the mother topology ${\cal A}_{212}$ where
$\WH\ell_1$ and $\WH\ell_2$ are mixed.

With above definition, we find the following reduction hinted by
unitarity method\footnote{For some topologies, such as ${\cal
A}_{112}$, since they are not detectable by our choice of unitarity
cuts, we can not find their coefficients.}
\bea
{\cal A}_{2 1 2}^{(a,b)} &  \to &  \sum_{\kappa_0,\kappa_1,\kappa_2}f_{212\to 202; \mu_1,...,\mu_{a};\nu_1,...,\nu_b}^{(a,b)} T_1^{\mu_1}...T_1^{\mu_a} T_2^{\nu_1}...
T_2^{\nu_b}{\cal B}_{202}[\kappa_0,\kappa_1,\kappa_2] \nn
& & +\sum_{\kappa_0,\kappa_1,\kappa_2}f_{212\to 212;
\mu_1,...,\mu_{a};\nu_1,...,\nu_b}^{(a,b)} T_1^{\mu_1}...T_1^{\mu_a}
T_2^{\nu_1}... T_2^{\nu_b}{\cal
B}_{212}[\kappa_0,\kappa_1,\kappa_2]~.~~~ \label{A212-final-2} \eea
However, before claiming ${\cal
B}_{212}[\kappa_0,\kappa_1,\kappa_2]$ are bases of the topology
${\cal A}_{212}$ studied in this paper, we need to notice that in
general $T_i$ could have four independent choices in 4D, i.e.,
$e_i$, $i=1,2,3,4$ as the momentum bases for Lorentz momenta. So if
bases have non-trivial dependence of $T_i$ in the numerator, we
should be careful to identify bases. This happens to topologies
${\cal A}_{213}$ and ${\cal A}_{313}$. However, for the current
topology ${\cal A}_{212}$, the bases ${\cal
B}_{212}[\kappa_0,\kappa_1,\kappa_2]$ are {\bf scalar bases}, i.e.,
the numerator of bases does not depend on any external momenta
$T_i$.

Now we count the number of integral bases. For pure 4D case, we can
take the limit $\mu_1^2,\mu_2^2,\mu_1\cdot \mu_2\to 0$, thus there
is only one scalar basis, with $\kappa_i=0$, $i=0,1,2$. In
\cite{Feng:2012bm} it is found that for planar double-triangle(i.e.,
the topology ${\cal A}_{212}$), the number of integrand bases is
$111$ under the renormalizable conditions in pure 4D. For general
$(4-2\eps)$-dimension, if we set constraint
$\sum_{i=0,1,2}\kappa_i\leq 3$(i.e., the sum of the power of
$\ell_1,\ell_2$ in the numerator is less than or equal to $6$) for
well-behaved quantum field theories, the number of integral bases is
$20$.

\section{The integral bases of ${\cal A}_{213}$ topology }

Encouraged by the results in previous section, in this section we
determine the integral bases of ${\cal A}_{213}$ topology. As it
will be shown shortly after, new features will appear.

\subsection{$\la_2$-integration for the case $n_2=3$}

For $n_2=3$ the general formula \eref{n1=2-la2} becomes(for
simplicity, we will drop "$\tau_i\to 0$" from now on)
\bea
\Delta{\cal A}_{2 1 3}^{(a,b)} & = & \int
d^{-2\eps} \mu_1 d^{-2\eps} \mu_2 \int \Spaa{\la_2|d\la_2}
\Spbb{\W\la_2|d\W\la_2}{  \Spab{\la_2|R_2|\W\la_2}^{b}
 \over  \Spab{\la_2|Q_{2}|\W\la_2}
 \Spab{\la_2|K_{L_2}|\W\la_2}^{b+1}} \nn & &
   \left\{{(-)^a t_2^{b-1}(K_{L_1}^2)^{a+b-1}\over a! t_2^a} \ln \left( { s+ t_1 t_2
   \over s-t_1 t_2}\right) {d^a\over
d\tau^a} {\Spab{\la_2|{\cal F}|\W\la_2}^a\over
\Spab{\la_2|K_{L_1}|\W\la_2}^a}\right.~~~\label{n2=2-la2-0}\\
& &\left. +\sum_{i=0}^{a-1} { (-)^{a-1} (K_{L_1}^2)^{a+b-1}
t_2^{b-1}  \over (a-i)a! t_2^{i}}{d^i\over d\tau^i}{d^{a-i}\over
d\tau_1^{a-i}} {\Spab{\la_2|{\cal G}_2|\W\la_2}^{a}+(-)^{a-1-i}
\Spab{\la_2|{\cal G}_1|\W\la_2}^{a}\over
\Spab{\la_2|K_{L_1}|\W\la_2}^{a}}\right\}~.~~~\nonumber \eea
Again, using $K_{L_1}=-K_{L_2}$ we can simplify the denominator.
There are also two parts we need to compute.

~\\ {\bf First part:} The first part of remaining integration can be
rewritten as
\bea
& & \int
d^{-2\eps} \mu_1 d^{-2\eps} \mu_2{(-)^{a+b+1} t_2^{b-1}(K_{L_1}^2)^{a+b-1}\over (a+b)! t_2^a} \ln \left( { (s+ t_1 t_2)\over (s-t_1 t_2)}\right) {d^a\over
d\tau^a}{d^b\over d\W\tau^b}\nn & & \int \Spaa{\la_2|d\la_2}
\Spbb{\W\la_2|d\W\la_2}{  \Spab{\la_2|\W\tau R_2+{\cal F}|\W\la_2}^{a+b}
 \over  \Spab{\la_2|Q_{2}|\W\la_2}
 \Spab{\la_2|K_{L_1}|\W\la_2}^{a+b+1}}~.~~~
\eea
The second line is the standard one-loop triangle  integration. The
one-loop triangle can be reduced to triangle part and bubble part,
thus they can be interpreted as contributions from topologies ${\cal
A}_{213}$ and ${\cal A}_{212}$.

~\\ {\bf Second part:} The second part can be written as
\bea
& & \int
d^{-2\eps} \mu_1 d^{-2\eps} \mu_2\sum_{i=0}^{a-1} { (-)^{a+b} (K_{L_1}^2)^{a+b-1} t_2^{b-1}  \over (a-i)(a+b)!
t_2^{i}}{d^i\over
d\tau^i}{d^{a-i}\over d\tau_1^{a-i}} {d^b \over d\W\tau^b}\nn
& &\int \Spaa{\la_2|d\la_2} \Spbb{\W\la_2|d\W\la_2}
{\Spab{\la_2|\W\tau R_2+{\cal G}_2|\W\la_2}^{a+b}+(-)^{a-1-i}
\Spab{\la_2|\W\tau R_2+ {\cal G}_1|\W\la_2}^{a+b}\over
\Spab{\la_2|Q_2|\W\la_2}\Spab{\la_2|K_{L_1}|\W\la_2}^{a+b+1}}~.~~~
\eea
The second line is again the standard triangle integration which
contain contributions from topologies ${\cal A}_{203}$ and ${\cal
A}_{202}$.

\subsection{Overview of results}

Collecting all results together, we get an expression of the form
\bea & & \Delta{\cal A}_{2 1 3}^{(a,b)}  =  \int d^{-2\eps} \mu_1
d^{-2\eps} \mu_2\left\{ f_{213\to 213}^{(a,b)} {\cal
S}_{213}+f_{213\to 212}^{(a,b)}{\cal S}_{212} + f_{213\to
203}^{(a,b)}{\cal S}_{203} +  f_{213\to 202}^{(a,b)}{\cal
S}_{202}\right\}~,~~~ \label{A213-final-0} \eea
where ${\cal S}_{202}$ and ${\cal S}_{212}$ have been defined in
\eref{212-sgn} and two new signatures are
\bea {\cal S}_{203}&=&{t_1\over 2\sqrt{(K_4\cdot
K_{L_1})^2-K_{L_1}^2 K_4^2}}\ln \left({K_4^2+K_4\cdot K_{L_1}-t_2
\sqrt{(K_4\cdot K_{L_1})^2-K_{L_1}^2 K_4^2}\over K_4^2+K_4\cdot
K_{L_1}+t_2 \sqrt{(K_4\cdot K_{L_1})^2-K_{L_1}^2 K_4^2}}
\right)~,~~~\nn
{\cal S}_{213}&=&{-\ln\left({s+t_1t_2\over s-t_1t_2}\right)\over
2t_2K_{L_1}^2\sqrt{(K_4\cdot K_{L_1})^2-K_{L_1}^2 K_4^2}}\ln
\left({K_4^2+K_4\cdot K_{L_1}-t_2 \sqrt{(K_4\cdot
K_{L_1})^2-K_{L_1}^2 K_4^2}\over K_4^2+K_4\cdot K_{L_1}+t_2
\sqrt{(K_4\cdot K_{L_1})^2-K_{L_1}^2 K_4^2}}
\right)~.~~~\label{313-sgn} \eea
There are a few remarks for expression \eref{A213-final-0}. Firstly
it is easy to see that the signature ${\cal S}_{203}$ is the direct
product of signatures of one-loop bubble and one-loop triangle.
Secondly there are two logarithms in the signature ${\cal S}_{213}$:
one depends on both $\mu_1,\mu_2$ and the other only depends on
$\mu_2$. Pictorially, the first logarithm is related to the mixed
propagator $(\ell_1+\ell_2)^2$ while the second logarithm is related
to the right hand side sub-triangle.

Thirdly, all dependence of $a,b$ are inside coefficients $f$ while
signatures are universal. However, unlike in the expression
\eref{A212-final-0} where coefficients $f$ are all rational
functions of external momenta and polynomials of $s, u_1, u_2$, here
we find that the coefficients $f$ are in general not polynomials of
$s,u_1, u_2$. In fact, factor $t_2=\sqrt{1-u_2}$ will appear in
denominators. Such behavior can not be explained by dimensional
shifted basis. Instead, we must regard it as the signature of new
integral basis. Because of such complexity, when talking about {\bf
the signature of a basis for ${\cal A}_{213}$ topology}, we should
treat all coefficients together in a  list $\{ f_{213\to
213}^{(a,b)},f_{213\to 212}^{(a,b)},f_{213\to 203}^{(a,b)},f_{213\to
202}^{(a,b)}\}$  as a single object. More explicitly we will write
the expression \eref{A213-final-0} as
\bea \Delta {\cal A}_{213}^{(a,b)} \equiv \{ f_{213\to
213}^{(a,b)},f_{213\to 212}^{(a,b)}, f_{213\to
203}^{(a,b)},f_{213\to 202}^{(a,b)}\}~.~~~\label{A213-list}\eea
The reduction of $\Delta {\cal A}_{213}^{(a,b)}$ is to write it as
the linear combination $\sum_{i} C_i \{ a_{i1}, a_{i2},
a_{i3},a_{i4} \}$ where $\{ a_{i1}, a_{i2}, a_{i3},a_{i4} \}$ is the
signature of $i$-th basis. In this notation, we can rewrite the
signatures of previously discussed bases as
\bea \Delta {\cal A}_{212}^{(0,0)}= \{0,1,0,0\}~~,~~\Delta {\cal
A}_{203}^{(0,0)} =\{0,0,1,0\}~~,~~\Delta{\cal
A}_{202}^{(0,0)}=\{0,0,0,1\}~.~~~\label{sig-list}\eea
Having above general remarks, now we present explicit results.

\subsection{The result of $a=0$}

We list results for $a=0$ with various $b$. Noticing that $a=0$
implies $f_{213\to 203}^{(0,b)}=0$ and $f_{213\to 202}^{(0,b)}=0$,
we will focus on the first two coefficients only.

~\\ {\bf The case $b=0$:} It is easy to see that
\bea \Delta {\cal A}_{213}^{(0,0)}=\{1, 0,0,0\}~.~~~
\label{213-a0b0-delta} \eea
Since it can not be written as the linear combination of three bases
in \eref{sig-list}, it must indicate a new integral basis.  In other
words, ${\cal A}_{213}^{(0,0)}$ is an integral basis with signature
\eref{213-a0b0-delta}.

~\\ {\bf The case $b=1$:} The result is
\bea \Delta {\cal A}_{213}^{(0,1)} = & &  \{{( K_4^2+K_4\cdot
K_{L_1}) (K_4\cdot T_2) K_{L_1}^2- K_4^2( K_4\cdot K_{L_1}+
K_{L_1}^2) (K_{L_1}\cdot T_2)\over  K_4^2 K_{L_1}^2- (K_4\cdot
K_{L_1})^2},\nn & & { (K_4\cdot K_{L_1})(T_2\cdot K_{L_1})-K_{L_1}^2
(K_4\cdot T_2) \over   K_4^2 K_{L_1}^2- (K_4\cdot
K_{L_1})^2},0,0\}~.~~~ \label{213-a0b1-delta} \eea
Thus, at least for our choice of unitarity cuts, $\Delta{\cal
A}_{213}^{(0,1)}$ can be written as the linear combination of
signatures   $\Delta{\cal A}_{213}^{(0,0)}$ and $\Delta{\cal
A}_{212}^{(0,0)}$ with rational coefficients of external momenta.

~\\ {\bf For other $b$'s:} We have calculated  cases $b=2$ and
$b=3$. Again we find that $\Delta {\cal A}_{213}^{(0,b)}$ can be
written as linear combinations of signatures  $\Delta{\cal
A}_{213}^{(0,0)}$ and $\Delta{\cal A}_{212}^{(0,0)}$, with
coefficients being rational functions of external momenta and
polynomials of $s, u_1, u_2$. The explicit expressions are too long
to write down here. When $s, u_1, u_2$ appear in the results, we
should include dimensional shifted bases too.

\subsection{The result of $a=1$}

In this case a non-trivial phenomenon appears, and we will show how
to explain it.

~\\ {\bf The case $b=0$:} Calculation yields
\bea f_{213\to 213}^{(1,0)} & = & {s\over t_2^2} f_{213\to
213;s^1}^{(1,0)} + f_{213\to 213;s^0}^{(1,0)}~,~~~\eea
where
\bean f_{213\to 213;s^1}^{(1,0)} & = & {( K_4^2+K_4\cdot K_{L_1})
[(K_4\cdot T_1) K_{L_1}^2-( K_4\cdot K_{L_1})( K_{L_1}\cdot T_1)]
\over   (K_4^2 K_{L_1}^2- (K_4\cdot K_{L_1})^2)}~,~~~\nn
f_{213\to 213;s^0}^{(1,0)} & = &  K_{L_1}\cdot T_1~.~~~\eean
Although the $s^0$-part can be explained by the signature $\Delta
{\cal A}_{213}^{(0,0)}$, the $s^1$-part with factor ${s\over t_2^2}$
can not because the appearance of $t_2^2=(1-u_2)$ in the
denominator. Thus factor ${s\over t_2^2}$ indicates a new integral
basis.

Besides $f_{213\to 213}^{(1,0)}$, other coefficients are given by
\bea f_{213\to 212}^{(1,0)}& = &{ s[- K_{L_1}^2 (K_4\cdot T_1)
+(K_4\cdot K_{L_1})(T_1\cdot K_{L_1})]\over t_2^2(-K_4^2 K_{L_1}^2+
(K_4\cdot K_{L_1})^2)}~,~~~\nn
 f_{213\to 203}^{(1,0)} & = & { -2(K_4^2+K_4\cdot K_{L_1})
 [ K_{L_1}^2 (K_4\cdot T_1)
-(K_4\cdot K_{L_1})(T_1\cdot K_{L_1})]\over t_2^2 K_{L_1}^2 (K_4^2
K_{L_1}^2- (K_4\cdot K_{L_1})^2)}~,~~~\nn
f_{213\to 202}^{(1,0)}& = &{ 2
 [- K_{L_1}^2 (K_4\cdot T_1)
+(K_4\cdot K_{L_1})(T_1\cdot K_{L_1})]\over t_2^2 K_{L_1}^2 (-K_4^2
K_{L_1}^2+ (K_4\cdot K_{L_1})^2)}~.~~~ \eea
Again, because of the factor ${1\over t_2^2}$, they can not be
explained by signatures \eref{sig-list}. Thus we have the first
non-trivial example of signatures where all four components are
non-zero
\bea \Delta{\cal A}_{213}^{(1,0)}= \{f_{213\to
213}^{(1,0)},f_{213\to 212}^{(1,0)},f_{213\to 203}^{(1,0)},f_{213\to
202}^{(1,0)}\}~.~~~\label{A213a1b0-sig}\eea

~\\ {\bf The case $b=1$:} All coefficients $\{f_{213\to
213}^{(1,1)},f_{213\to 212}^{(1,1)},f_{213\to 203}^{(1,1)},f_{213\to
202}^{(1,1)}\}$ have ${1\over t_2^2}$ dependence. However, all these
${1\over t_2^2}$ factors can be absorbed into $\Delta {\cal
A}_{213}^{(1,0)}$. More explicitly, we found the following
decomposition
\bea \Delta {\cal A}_{213}^{(1,1)} & = & a_{11\to 00}\Delta {\cal
A}_{213}^{(0,0)} + a_{11\to 10} \Delta {\cal
A}_{213}^{(1,0)}+b_{11\to 00}\Delta {\cal A}_{212}^{(0,0)}+ d_{11\to
00}\Delta {\cal A}_{202}^{(0,0)}~,~~~\label{A213-a1b1-delta} \eea
where
\bean a_{11\to 10} & = & { (1-u_2)K_{L_1}^2 ( (K_4\cdot
K_{L_1})^2-K_4^2 K_{L_1}^2) \Sigma_1+(K_4^2+ K_4\cdot
K_{L_1})\Sigma_2 \over 2 (K_4^2+ K_4\cdot K_{L_1}) ( (K_4\cdot
K_{L_1})^2-K_4^2 K_{L_1}^2)[(K_4\cdot T_1)K_{L_1}^2 -(K_4\cdot
K_{L_1})(T_1\cdot K_{L_1})]}~,~~~\nn
a_{11\to 00} & = & { (K_{L_1}\cdot T_1) [-(K_4\cdot T_2)K_{L_1}^2
(K_4^2+ K_4\cdot K_{L_1})+ K_4^2 (K_4\cdot K_{L_1}+K_{L_1}^2)(
K_{L_1}\cdot T_2)] \over ( (K_4\cdot K_{L_1})^2-K_4^2
K_{L_1}^2)}-(K_{L_1}\cdot T_1)a_{11\to 10}~,~~~\eean
\bean b_{11\to 00} & = & {1\over 2 (K_4\cdot K_{L_1}+K_4^2) (
-(K_4\cdot K_{L_1})^2+K_4^2 K_{L_1}^2)}\left\{ 2 (K_4\cdot
K_{L_1}+K_4^2) K_{L_1}\cdot T_1\right. \nn
& &  ( -K_{L_1}^2( K_4\cdot T_2) +(K_4\cdot K_{L_1})( K_{L_1}\cdot
T_2))+ s K_{L_1}^2 ( K_{L_1}\cdot T_1 ( K_4\cdot K_{L_1} (K_4\cdot
T_2) -K_{4}^2 (K_{L_1}\cdot T_2))\nn & & \left.+K_4\cdot T_1 (
-K_{L_1}^2 (K_4\cdot T_2) + K_4\cdot K_{L_1} (K_{L_1}\cdot T_2)+
T_1\cdot T_2 ( -(K_4\cdot K_{L_1})^2+K_4^2
K_{L_1}^2))\right\}~,~~~\nn
d_{11\to 00} & = & {K_{L_1}\cdot T_1 (K_4\cdot K_{L_1}(K_4\cdot T_2)
-K_4^2 (K_{L_1}\cdot T_2))+ K_4\cdot T_1 (- K_{L_1}^2 (K_4\cdot T_2)
+K_4\cdot K_{L_1} (K_{L_1}\cdot T_2)) \over (K_4\cdot K_{L_1}+K_4^2)
( -(K_4\cdot K_{L_1})^2+K_4^2 K_{L_1}^2)}\nn & & + {T_1\cdot T_2
\over (K_4\cdot K_{L_1}+K_4^2)}~,~~~ \eean
with
\bean \Sigma_1&=&(K_{L_1}\cdot T_1)( -(K_4\cdot K_{L_1}) (K_4\cdot
T_2) +K_4^2 (K_{L_1}\cdot T_2))+ K_4\cdot T_1 ( (K_4\cdot
T_2)K_{L_1}^2 \nn & &-(K_4\cdot K_{L_1})(T_2\cdot K_{L_1})) + (
(K_4\cdot K_{L_1})^2-K_4^2 K_{L_1}^2) T_1\cdot T_2~,~~~\nn
\Sigma_2 & = & (K_4^2)^2 K_{L_1}^2( -(K_{L_1}\cdot T_1)(K_{L_1}\cdot T_2)
+K_{L_1}^2 (T_1\cdot T_2)) + (K_4\cdot K_{L_1}) K_{L_1}^2 [ K_4\cdot T_1
(-3 (K_4\cdot T_2) K_{L_1}^2\nn & & + (K_4\cdot K_{L_1}) (K_{L_1}\cdot T_2))
+ K_4\cdot K_{L_1} (3 (K_4\cdot T_2) (K_{L_1}\cdot T_1)
-(K_4\cdot K_{L_1}) (T_1\cdot T_2))]\nn & & + K_4^2( K_4\cdot T_1 K_{L_1}^2
(-3 K_4\cdot T_2 K_{L_1}^2+(3 K_4\cdot K_{L_1} +2 K_{L_1}^2) K_{L_1}\cdot T_2)
+ (K_4\cdot K_{L_1})\nn & &  ( (K_{L_1}\cdot T_1)(3 K_{L_1}^2 (K_4\cdot T_2-K_{L_1}\cdot T_2)
- 2 (K_4\cdot K_{L_1}) (K_{L_1}\cdot T_2))\nn
& &+ K_{L_1}^2(-K_4\cdot K_{L_1}+K_{L_1}^2) T_1\cdot T_2)~.~~~ \eean
Since above four coefficients are rational functions of  external
momenta and polynomials of $u_2$, we can claim that ${\cal
A}_{213}^{(1,1)}$ is not a basis at least for our choice of
unitarity cuts.

There are some details we want to remark. The coefficient $a_{11\to
00}$ is a polynomial of $T_1$ and $T_2$ with degree one while
coefficient $a_{11\to 10}$ is a polynomial of $T_2$ with degree one
but rational function of $T_1$. More  accurately, both the
denominator and the numerator of $a_{11\to 10}$ are polynomials of
$T_1$ with degree one.  It is against the intuition since $T_1$
should not appear in the denominator. However, this subtlety is
resolved if one notice that the first component $f_{213\to
213;s^1}^{(1,0)}$ of $\Delta {\cal A}_{213}^{(1,0)}$ contains
exactly the same factor $ [-(K_4\cdot T_1) K_{L_1}^2+( K_4\cdot
K_{L_1})( K_{L_1}\cdot T_1)]$ in its numerator, so it cancels the
same factor in denominator of $a_{11\to 10}$.

~\\ {\bf The case $b=2$:} The whole expression is too long to write
down, thus we present only the general feature. Again although all
coefficients contain factor ${1\over t_2^2}$,  the whole result can
be expanded like the one \eref{A213-a1b1-delta} with coefficients as
rational functions of external momenta and polynomials of $s,u_1,
u_2$. Thus ${\cal A}_{213}^{(1,2)}$ is not a new integral basis.

\subsection{The result of $a=2$}

We will encounter similar phenomenon as in the case $a=1$. To get
rid of tedious expressions, we will present only the main features.

~\\ {\bf The case $b=0$:} The coefficient $f_{213\to 213}^{(2,0)}$ has the following form
\bea
f_{213\to 213}^{(2,0)}= {s^2\over t_2^4} g_{0;1}+ {s^2\over t_2^2} g_{0;2}+
{s\over t_2^2} g_{0;3}+{1\over t_2^2} g_{0;4}+ g_{0;5}~,~~~
\eea
where $g_{0;i}$'s are polynomials of $u_1$ and rational functions of
external momenta(similar for all other coefficients such as $h,i,j$
in this subsection). The appearance of $g_{0;1}$-part and
$g_{0;4}$-part can not be counted by signatures $\Delta{\cal
A}_{213}^{(0,0)}$ and $\Delta{\cal A}_{213}^{(1,0)}$, thus we should
take ${\cal A}_{213}^{(2,0)}$ as a new integral basis. For other
coefficients, we have
\bea
f_{213\to 212}^{(2,0)} & = & {s^2\over t_2^4} h_{0;1}+ {s\over t_2^2} h_{0;2}+
{1\over t_2^2}  h_{0;3}~,~~~ \nn
f_{213\to 203}^{(2,0)} & = & {s\over t_2^4} i_{0;1}+ {s\over t_2^2} i_{0;2}+
{1\over t_2^2}  i_{0;3}~,~~~\nn
f_{213\to 202}^{(2,0)}& = & {s\over t_2^4} j_{0;1}+
{1\over t_2^2}  j_{0;2}~.~~~
\eea
The signature of the new basis can be represented by
\bea \Delta {\cal A}_{213}^{(2,0)}= \{ f_{213\to 213}^{(2,0)},
f_{213\to 212}^{(2,0)}, f_{213\to 203}^{(2,0)}, f_{213\to
202}^{(2,0)}\}~.~~~\eea

~\\ {\bf The case of $b=1$:} The behavior of various coefficients are
\bea
f_{213\to 213}^{(2,1)} &= & {s^2\over t_2^4} g_{1;1}+ {s^2( g_{1;2;0}+t_2^2 g_{1;2;1})\over t_2^2} +
{s (g_{1;3;0}+t_2^2 g_{1;3;1})\over t_2^2} +{1\over t_2^2} g_{1;4}+ g_{1;5}~,~~~\nn
f_{213\to 212}^{(2,1)} & = & {s^2 (h_{1;1;0}+t_2^2 h_{1;1;1})\over t_2^4}+ {s\over t_2^2} h_{1;2}+
{1\over t_2^2}  (h_{1;3;0}+t_2^2 h_{1;3;1})~,~~~ \nn
f_{213\to 203}^{(2,1)} & = & {s\over t_2^4} i_{1;1}+ {s\over t_2^2} i_{1;2}+
{1\over t_2^2} ( i_{1;3;0}+t_2^2 i_{1;3;1})~,~~~\nn
f_{213\to 202}^{(2,1)}& = & {s(j_{1;1;0}+t_2^2 j_{1;1;1})\over t_2^4} +
{1\over t_2^2}  j_{1;2}~,~~~\eea
where the integer $n$ in $g_{1;m;n}$ denotes the power of $t_2^2$, and similar for $h,~i,~j$.

We found the following  expansion
\bea \Delta {\cal A}_{213}^{(2,1)} & = & a_{21\to 20}\Delta {\cal
A}_{213}^{(2,0)} + a_{21\to 10} \Delta {\cal A}_{213}^{(1,0)}+
a_{21\to 00}\Delta {\cal A}_{213}^{(0,0)}+b_{21\to 00}\Delta {\cal
A}_{212}^{(0,0)}\nn & & + c_{21\to 00}\Delta {\cal A}_{203}^{(0,0)}+
d_{21\to 00}\Delta {\cal
A}_{202}^{(0,0)}~,~~~\label{A213-a2b1-delta} \eea
where coefficients are rational functions of external momenta and
polynomials of $s,u_1, u_2$. Thus ${\cal A}_{213}^{(2,1)}$ is not a
new integral basis.

\subsection{Classification of integral bases}

With above results, we can classify the integral bases of ${\cal
A}_{213}$ topology. Having shown that coefficients such as $a_{21\to
00}$ are polynomials of $\mu_1\cdot \mu_2, \mu_1^2,
\mu_2^2$ and rational functions of external momenta, we can expand
them, for example
\bea a_{21\to 00} & = & \sum_{\kappa_0,\kappa_1,\kappa_2} a_{21\to
00}^{(a,b)} (\mu_1^2)^{\kappa_1} (\mu_2^2)^{\kappa_2} (\mu_1\cdot
\mu_2)^{\kappa_0}~,~~~ \eea
where the tensor coefficients $a$ are rational functions of external
momenta. This expansion leads us to define the following dimensional
shifted bases
\bea {\cal B}_{213;a}[\kappa_0,\kappa_1,\kappa_2;T_1] & \equiv &
\int d^{4-2\eps}\WH\ell_1\int d^{4-2\eps}\WH\ell_2
{(\mu_1^2)^{\kappa_1} (\mu_2^2)^{\kappa_2} (\mu_1\cdot
\mu_2)^{\kappa_0} (\WH \ell_1\cdot T_1)^a \over \WH\ell_1^2
(\WH\ell_1-K_{L_1})^2 \WH\ell_2^2 (\WH \ell_2-K_4)^2
(\WH\ell_2+K_{L_1})^2(\WH\ell_1+\WH\ell_2)^2}~.~~~\label{B213-dim-basis-0}
\eea

Unlike the scalar basis ${\cal B}_{212}[\kappa_0,\kappa_1,\kappa_2]
$ for ${\cal A}_{212}$ topology, the basis ${\cal
B}_{213;a}[\kappa_0,\kappa_1,\kappa_2] $ depends on $T_1$
explicitly. Since $T_1$ is a 4-dimensional Lorentz vector, there are
four independent choices and we need to clarify if different choice
of $T_1$ gives new independent basis.

To discuss this problem we expand $T_1=\sum_{i=1}^4 x_i e_i$. The momentum
bases $e_i$ are constructed as follows. Using $K_4, K_{L_1}$ we
can construct two null momenta $P_{i}= K_4+w_i K_{L_1}$ with $w_i=
{-K_4\cdot K_{L_1}\pm \sqrt{(K_{L_1}\cdot K_4)^2- K_4^2
K_{L_1}^2}\over K_{L_1}^2}$, thus the momentum bases can be taken as
\bea e_1=
K_4~~,~~e_2=K_{L_1}~~,~~e_3=\ket{P_1}\bket{P_2}~,~~e_4=\ket{P_2}\bket{P_1}~.~~~\label{e-basis}
\eea
~\\{\bf The case $a=0$:}
For $a=0$, since $T_1$ does not appear, only {\bf scalar basis}
exist. Thus the independent integral bases are ${\cal
B}_{213;0}[\kappa_0,\kappa_1,\kappa_2] $.
~\\{\bf The case $a=1$:}
We set $T_1=e_i$ for $i=1,2,3,4$ in
the expressions $f_{213\to 213}^{(1,0)}$, $f_{213\to 212}^{(1,0)}$,
$f_{213\to 203}^{(1,0)}$, $f_{213\to 202}^{(1,0)}$, and found that:
\begin{itemize}

\item (1) For $T_1=e_3$ or $T_1=e_4$ we have
\bea \{ f_{213\to 213}^{(1,0)},f_{213\to 212}^{(1,0)},f_{213\to 203}^{(1,0)},f_{213\to 202}^{(1,0)}\}
=\{0,0,0,0\}~.~~~ \eea
It can be shown that  $T_1=e_{3,4}$ are spurious and the
integrations are zero.

\item (2) For $T_1=K_{L_1}$, we find
\bea \{ f_{213\to 213}^{(1,0)},f_{213\to 212}^{(1,0)},f_{213\to 203}^{(1,0)},f_{213\to 202}^{(1,0)}\}
=\{K_{L_1}^2,0,0,0\}~.~~~ \eea
It is, in fact, equivalent to the basis ${\cal
B}_{213;0}[\kappa_0,\kappa_1,\kappa_2] $ and does not give new
integral basis.

\item (3) For $T_1=K_4$, we find
\bea
& & \{ f_{213\to 213}^{(1,0)},f_{213\to 212}^{(1,0)},f_{213\to 203}^{(1,0)},f_{213\to 202}^{(1,0)}\}
\nn
& = &\{{-s(K_4^2+ K_4\cdot K_{L_1})+t_2^2 K_4\cdot K_{L_1}\over
t_2^2 },{s\over t_2^2}, -{2(K_4^2+ K_4\cdot K_{L_1})\over t_2^2
K_{L_1}^2} ,{2\over t_2^2 K_{L_1}^2}\}~,~~~\label{A213-T1=K4} \eea
which is the true new integral basis.

\end{itemize}
~\\{\sl Conclusion:} For $a=1$, the integral basis is given by
${\cal B}_{213;1}[\kappa_0,\kappa_1,\kappa_2;K_4]$
~\\{\bf The case $a=2$:}
There are ten possible combinations $(\WH \ell_1\cdot e_i)(\WH \ell_1\cdot e_j)$. With the
explicit result we found that
\begin{itemize}

\item(1) For the following six combinations
\bea
(e_i,e_j)=(e_3,e_3)~,~(e_4,e_4)~,~(e_1,e_3)~,~(e_2,e_3)~,~(e_1,e_4)~,~(e_2,e_4)~,~~~
\eea
the coefficients are $\{0,0,0,0\}$.
In fact, integrations for these six cases are zero.

\item(2) For $(e_i,e_j)=(e_2,e_2)$ the list of coefficients is $
\{ 2 (K_{L_1}^2)^2,0,0,0\}$. It is equivalent to the basis ${\cal
B}_{213;0}[\kappa_0,\kappa_1,\kappa_2] $. Therefore it does not give
new integral basis.

\item(3) For  $(e_i,e_j)=(e_1,e_2)$ the list of coefficients is
\bea
& &
2K_{L_1}^2\{{-s(K_4^2+ K_4\cdot K_{L_1})+t_2^2 K_4\cdot K_{L_1}\over t_2^2 },{s\over t_2^2},
-{2(K_4^2+ K_4\cdot K_{L_1})\over t_2^2 K_{L_1}^2} ,{2\over t_2^2 K_{L_1}^2}\}~,~~~
\eea
which is proportional to \eref{A213-T1=K4} by a factor $2 K_{L_1}^2$. Therefore it can be reduced to the
basis ${\cal B}_{213;1}[\kappa_0,\kappa_1,\kappa_2] $, and dose not give a new integral basis.

\item(4) For  $(e_i,e_j)=(e_1,e_1)$ and $(e_i,e_j)=(e_3,e_4)$ the list is non-trivial. However,
it can be checked that
\bea
& &  \{ f_{213\to 213}^{(2,0)},f_{213\to 212}^{(2,0)},f_{213\to 203}^{(2,0)},f_{213\to 202}^{(2,0)}\}|_{(e_i,e_j)=(e_1,e_1)}\nn
& = &
\{ f_{213\to 213}^{(2,0)},f_{213\to 212}^{(2,0)},f_{213\to 203}^{(2,0)},f_{213\to 202}^{(2,0)}\}|_{(e_i,e_j)=(e_3,e_4)}\nn
& & + 2 (K_4\cdot K_{L_1})\{ f_{213\to 213}^{(1,0)},f_{213\to 212}^{(1,0)},f_{213\to 203}^{(1,0)},f_{213\to 202}^{(1,0)}\} \nn
& & + ( (t_1^2-1) (K_4\cdot K_{L_1})^2- t_1^2 K_4^2 K_{L_1}^2)\{ f_{213\to 213}^{(0,0)},f_{213\to 212}^{(0,0)},
f_{213\to 203}^{(0,0)},f_{213\to 202}^{(0,0)}\}~.~~~
\eea
Thus we can take either one(but only one of them) as the
integral basis. We choose the combination $(e_i,e_j)=(e_1,e_1)$
to be a new integral basis.

\end{itemize}

~\\{\sl Conclusion:} For $a=2$, the integral basis can be chosen as
${\cal B}_{213;2}[\kappa_0,\kappa_1,\kappa_2;K_4]$

~\\{\bf For general $a$:} Although we have not done explicit
calculations for $a\geq 3$, we expect for each $a$ there is a new
integral bases ${\cal B}_{213;a}[\kappa_0,\kappa_1,\kappa_2;K_4]$.

~\\ {\bf The number of integral bases:} To finish this section, let
us count the number of integral basis. For pure 4D, we just need to
set $\mu_1\cdot \mu_2,\mu_1^2, \mu_2^2$ to zero. In this case, the
factor ${1\over t_2^n}\to 1$. In other words, there is only one
integral bases
\bea \int d^{4-2\eps}\WH\ell_1\int d^{4-2\eps}\WH\ell_2 {1 \over
\WH\ell_1^2 (\WH\ell_1-K_{L_1})^2 \WH\ell_2^2 (\WH \ell_2-K_4)^2
(\WH\ell_2+K_{L_1})^2(\WH\ell_1+\WH\ell_2)^2 }~.~~~\eea
It is useful to compare it with about $70$ integrand bases found in
\cite{Feng:2012bm} under renormalizable conditions.

For general $(4-2\eps)$-dimension, renormalizable conditions can be
roughly given by $2\kappa_1+a\leq 3$, $\kappa_2\leq 2$. Under these
two conditions, we find $48$ integral bases.

\section{The integral basis of ${\cal A}_{313}$ topology}

In this section we turn to the topology ${\cal A}_{313}$. This
topology has been extensively studied by various methods, such as
IBP method\cite{Gluza:2010ws} and maximum unitarity cut
method\cite{Kosower:2011ty,Larsen:2012sx,CaronHuot:2012ab,
Johansson:2012zv,Johansson:2012sf,Sogaard:2013yga,Johansson:2013sda},
and its integral bases have been determined\cite{Gluza:2010ws}. To
determine the integral bases using our method, we need to integrate
the following expression
\bea \Delta{\cal A}_{3 1 3}^{(a,b)}&=&\int d^{-2\eps}\mu_1
d^{-2\eps}\mu_2\int d^4 \W\ell_2~
\delta(\W\ell_2^2-\mu_2^2)\delta(K_{L_2}^2-2K_{L_2}\cdot \W\ell_2){
(2\W\ell_2\cdot T_2)^b \over((\W\ell_2-K_{4})^2-\mu_2^2)}\nn
& &\int\Spaa{\la_1|d\la_1}\Spbb{\W\la_1|d\W\la_1} { - ((1-2z_1)
K_{L_1}^2)^{a-1}\over \Spab{\la_1|K_{L_1}|\W\la_1}^{a}}{
\Spab{\la_1|R_1|\W\la_1}^a\over \Spab{\la_1|W_1|\W\la_1}
\Spab{\la_1|Q_{1}|\W\la_1}}~~~\label{redo-L1-delta-n1=3}~,~~~\eea
with $K_{L_1}=K_1+K_2$. For general situation, the integration is
very complicated and we postpone it to future study. In this paper,
we take the following simplification. Firstly we take all out-going
momenta $K_i^2=0\ (i=1,..,4)$(unlike the topologies ${\cal A}_{212}$
and ${\cal A}_{213}$ where $K_i$ can be massive). Secondly, based on
the known results of integral bases, we focus on the specific case
$a=0$ and $T_2=K_1$.

In order to make expressions compact we define some new parameters
as\footnote{It is worth to notice that $s$ in this section is
different from $s$ in \eref{combi} of section 3. }
\begin{eqnarray}
s\equiv s_{12}~~,~~ m\equiv\frac{s_{14}-s_{13}}{s_{12}}~~,~~
\chi\equiv\frac{s_{14}}{s_{12}}=\frac{m-1}{2}~.~~~
\end{eqnarray}
For physical unitarity cut, momentum configuration requires
$s_{12}>0$, $s_{13}<0$ and $s_{14}<0$. So we have
\begin{eqnarray}
-1<m<1~~,~~ -1<\chi<0
\end{eqnarray}
by momentum conservation $s_{12}+s_{13}+s_{14}=0$. Furthermore, we
define the regularization parameters $\gamma_i$ as
\begin{eqnarray}
\gamma\equiv\frac{1+\nu_1\cdot\nu_2}{\sqrt{1-\nu^2_1}\sqrt{1-\nu^2_2}}~~,~~
\gamma_i\equiv\frac{1}{\sqrt{1-\nu^2_i}}~~,~~
i=1,2~,~~~\label{gamma-i}
\end{eqnarray}
where the dimensionless extra-dimensional vector $\nu_i$ is defined
as  $\nu_i\equiv2\mu_i/\sqrt{s}~,~i=1,2$.

Under the simplification $a=0$, the integration over $\la_1$-part is
trivial.  Using \eref{Stan-box-box} in Appendix \ref{B} we can get
\bea \Delta{\cal A}_{3 1 3}^{(0,b)} & = & \int d\mu_i
\int\Spaa{\la_2|d\la_2}\Spbb{\W\la_2|d\W\la_2}{\gamma_1\over
s^2}\Big(-{s\over
\gamma_2}\Big)^{b-1}{\Spab{\la_2|R_2|\W\la_2}^b\over
\Spab{\la_2|Q_2|\W\la_2}\Spab{\la_2|K_{L_1}|\W\la_2}^{b+1}} \nn & &
{1\over \sqrt{{\Spab{\la_2|\W K_1|\W\la_2}^2\over
\Spab{\la_2|K_{L_1}|\W\la_2}^2}- {\b^2\over 4} }}
\ln\left({\Spab{\la_2|\W K_1|\W\la_2}\over
\Spab{\la_2|K_{L_1}|\W\la_2}}+\sqrt{{\Spab{\la_2|\W
K_1|\W\la_2}^2\over \Spab{\la_2|K_{L_1}|\W\la_2}^2}- {\b^2\over 4} }
\over {\Spab{\la_2|\W K_1|\W\la_2}\over
\Spab{\la_2|K_{L_1}|\W\la_2}}-\sqrt{{\Spab{\la_2|\W
K_1|\W\la_2}^2\over \Spab{\la_2|K_{L_1}|\W\la_2}^2}- {\b^2\over 4}
}\right)~,~~~\label{A-exp}\eea
where
\bean \b^2=(\gamma^2-1)(\gamma^2-1)~.~~~\eean
An important feature is that the signature after $\la_1$-integration
depends on $\ell_2$ explicitly, which is different from the
signature in \eref{3to3-sign}. Because of this, the integration over
$\la_2$ becomes very complicated. One way to overcome is to use
\bea {1\over b} \log {a+b\over a-b}=\int_0^1 dx\left( {1\over a+x
b}+{1\over a-x b}\right)=\int_0^1 dx {2a\over a^2-x^2 b^2 }
~.~~~\label{Useful-1}\eea
Thus the logarithmic part in \eref{A-exp} becomes rational function
of $\ell_2$ and we can use the same strategy as in previous
sections. However, for the current simple situation,
 we can use another method. After  expanding the spinor variables  as
\bea
\ket{\la_2}=\ket{k_2}+z\ket{k_1}~~,~~\bket{\W\la_2}=\bket{k_2}+\O
z\bket{k_1}~~,~~\Spaa{\la_2|d\la_2}\Spbb{\W\la_2|d\W\la_2}=-s dz d\O
z~,~~~\eea
the integration becomes an integration over complex plane
\bea \Delta{\cal A}_{3 1 3}^{(0,b)} =  \int d\mu_i \int
|dzd\bar{z}|(\bullet)=\int d\mu_i \int _0^{+\infty}r
dr\int_0^{2\pi}d\theta (\bullet)~~,~~z=r e^{i\theta}~.~~~ \eea

~\\ {\bf  $\theta$-integration:} The $\theta$-dependent part of
\eref{A-exp} is given by
\bea \int_0^{2\pi} d\theta{
\Big(\Spab{K_2|R_2|K_2}+r^2\Spab{K_1|R_2|K_1}+r
e^{i\theta}\Spab{K_1|R_2|K_2} +r
e^{-i\theta}\Spab{K_2|R_2|K_1}\Big)^b\over (s_{24}-\W t_2 s_{12})+
r^2(s_{14}-\W t_2 s_{12}) +r e^{i\theta} \Spab{K_1|K_4|K_2}+r
e^{-i\theta}\Spab{K_2|K_4|K_1}}~,~~~\eea
with $\W t_2={\gamma_2-1\over 2}$. Setting $x=e^{i\theta}$ the
integral becomes a circle contour integration with radius one
\bea \oint_{|x|=1}dx {\Big(x\Spab{K_2|R_2|K_2}+x
r^2\Spab{K_1|R_2|K_1}+r x^2\Spab{K_1|R_2|K_2}+r
\Spab{K_2|R_2|K_1}\Big)^b \over i x^b \Big(x(s_{24}-\W t_2s_{12})+
xr^2(s_{14}-\W t_2 s_{12})
+rx^2\Spab{K_1|K_4|K_2}+r\Spab{K_2|K_4|K_1}\Big)}~.~~~\label{theta-1}
\eea
There are three poles in total. The first one is $x=0$ when $b\neq
0$ for general $R_2$. The other two are  roots of the quadratic
polynomial in denominator
\bea x_{1,2}={-\Big(s_{24}-s_{14}+(r^2+1)(s_{14}-\W t_2s_{12})\Big)
\pm\sqrt{\Delta} \over 2r\Spab{K_1|K_4|K_2}}~,~~~\eea
where
\bea \Delta=\Big(-s_{12}+(r^2+1)(s_{14}-\W t_2
s_{12})\Big)^2+4s_{12}s_{14}(r^2+1)(1+\W t_2)~. \eea
It is easy to check that $|x_1 x_2|=1$. Thus one root is inside the
integration contour and the other is outside. The kinematic
conditions $s_{12}>0, s_{24}<0, s_{14}<0$ ensure that $x_1$ is the
one inside. The residue at the pole $x_1$ is
\bea {1\over i\sqrt{\Delta}}
\Big(\Spab{K_2|R_2|K_2}+r^2\Spab{K_1|R_2|K_1}+r
x\Spab{K_1|R_2|K_2}+r x^{-1}
\Spab{K_2|R_2|K_1}\Big)^b_{x=x_1}~.~~~\eea

~\\ {\bf The case $(T_2=K_1)$:} Under our simplification, we set
$T_2=K_1$, thus $\Spab{K_1|R_2|K_2}=0$ and $\Spab{K_2|R_2|K_1}=0$.
Because of this, there is no pole at $x=0$ in \eref{theta-1}. Thus
after the $\theta$-integration, \eref{A-exp} is reduced to
\bea \Delta{\cal A}^{(0,b)}_{313} &=&\int
d\mu_i~~\frac{\gamma_1}{2s^2}(-\frac{s}{\gamma_2})^{b-1}
\int_0^{+\infty}dr^2 {1\over \sqrt{\left({\alpha-1\over 2}+{1\over
(1+r^2)}\right)^2-{\b^2\over 4} }}\nn & &
\left(\ln{\left({\alpha-1\over2}+{1\over 1+r^2}\right)
+\sqrt{\left({\alpha-1\over2}+{1\over 1+r^2}\right)^2-{\b^2\over 4}
}\over \left({\alpha-1\over 2}+{1\over 1+r^2}\right)-
\sqrt{\left({\alpha-1\over 2}+{1\over 1+r^2}\right)^2-{\b^2\over 4}
}} \right) {1\over 1+r^2}\left({\gamma_2-1\over 2}+ {1\over 1+r^2}
\right)^b\nn & &   {1\over \sqrt{\Big((r^2+1) (\chi-{\gamma_2-1\over
2})-1\Big)^2 +4(r^2+1) \chi(1+{\gamma_2-1\over 2})}}~,~~~ \eea
in which
\begin{eqnarray}
\qquad\alpha=\gamma\gamma_1~~,~~
\beta=\sqrt{(\gamma^2-1)(\gamma^2_1-1)}~.~~~ \nonumber
\end{eqnarray}
Defining $u= {1-r^2\over 1+r^2}$ we arrive
\begin{eqnarray}\label{new expression a=0}
\Delta{\cal A}^{(0,b)}_{313} &=&\int
d\mu_i~~\frac{\gamma_1}{2s^2}(-\frac{s}{2\gamma_2})^{b-1}\int^{+1}_{-1}du
\frac{(u+\gamma_2)^b}{\sqrt{(u+m\gamma_2)^2+(1-m^2)(\gamma^2_2-1)}} \nonumber \\
&&\frac{1}{\sqrt{(u+\alpha)^2-\beta^2}}\ln\frac{(u+\alpha)+\sqrt{(u+\alpha)^2
-\beta^2}}{(u+\alpha)-\sqrt{(u+\alpha)^2-\beta^2}}~.~~~\label{A-exp-3}
\end{eqnarray}
An important observation from  \eref{A-exp-3} is that ${\cal
D}^{(0,b)}_{313}/(\gamma_1\gamma_2)$ has the symmetry
$\gamma\leftrightarrow \gamma_1$ as well as the symmetry
$\gamma_2\leftrightarrow \gamma_1$ for $b=0$ by the topology.

Since we use the dimensional shifted bases, the $\mu_i$ part is kept
and we will focus on  ${\cal D}_{313}^{(0,b)}$ after the
$u$-integration, i.e.,
\bea \Delta{\cal A}_{313}^{(0,b)}\equiv\int d\mu_i~{\cal
D}_{313}^{(0,b)}~.~~~\eea
We found it hard to integrate over $u$ and get analytic results.
However, in the general $(4-2\eps)$-dimensional framework, we can
treat $\mu_i^2$ and $\mu_1\cdot \mu_2$ as small parameters and take
series expansion around $\mu_i^2\to 0$. It is equivalent to taking
the series expansion around $\gamma_i\to 1$. The details of
calculation can be found in Appendix C. Up to the leading order,
result for $a=0,b=0$  is given by
\begin{eqnarray}
{\cal D}^{(0,0)}_{313}=\frac{1}{s^3\ 2\chi}\Bigg[
&&\ln\Big(\frac{-2\chi}{\gamma-1}\Big)\ln\Big(\frac{-2\chi}{\gamma_1-1}\Big)+\ln\Big(\frac{-2\chi}{\gamma-1}\Big)\ln\Big(\frac{-2\chi}{\gamma_2-1}\Big)
+\ln\Big(\frac{-2\chi}{\gamma_1-1}\Big)\ln\Big(\frac{-2\chi}{\gamma_2-1}\Big) \nonumber \\
&&+2\
\textsf{Li}_2(1+\chi)-\frac{\pi^2}{3}\Bigg]~.~~~\label{b=0-result}
\end{eqnarray}
An important check for the result \eref{b=0-result} is that it has
the $S_3$ permutation symmetry among $\gamma_1, \gamma_2,\gamma$.
The terms $\ln(-\chi)$ and $\textsf{Li}_2(1+\chi)$ do not show up
for topologies ${\cal A}_{212}$ and ${\cal A}_{213}$, thus they
belong to the signature of ${\cal A}_{313}$. For $b=1$, the result
is
\begin{eqnarray}
{\cal D}^{(0,1)}_{313}=\chi\ s{\cal D}^{(0,0)}_{313}+{\cal
D}^{(0,0)}_{312} -\frac{1}{s^2}\ln\Big(\frac{-2\chi}{\gamma-1}\Big)
\ln\Big(\frac{-2\chi}{\gamma_1-1}\Big)~.~~~\label{b=1-result}
\end{eqnarray}
The extra term $-\frac{1}{s^2}\ln\Big(\frac{-2\chi}{\gamma-1}\Big)
\ln\Big(\frac{-2\chi}{\gamma_1-1}\Big)$ in \eref{b=1-result}
indicates that comparing to ${\cal D}^{(0,0)}_{313}$, ${\cal
D}^{(0,1)}_{313}$ should be taken as a new integral basis. For
$b=2$, the result is
\bea {\cal D}^{(0,2)}_{313}&=&\chi\ s{\cal
D}^{(0,1)}_{313}+\frac{2\chi+1}{s}{\cal D}^{(0,0)}_{202}
-\frac{2\chi+1}{2}{\cal D}^{(0,0)}_{212}-\frac{2\chi+1}{2}{\cal
D}^{(0,0)}_{302} -\frac{2\chi}{s}\ln(-\chi)~.~~~\label{b=2-result}
\eea
For this result, there are a few things we want to discuss. Firstly,
the same coefficient $-\frac{2\chi+1}{2}$ appears for ${\cal
D}^{(0,0)}_{212}$ and ${\cal D}^{(0,0)}_{302}$, which is the
consequence of symmetry $\gamma\leftrightarrow \gamma_1$ in
\eref{A-exp-3}. Secondly, the appearance of term $\ln(-\chi)$ is
quite intriguing. There are several possible interpretations:
\begin{itemize}

\item Under the general
$(4-2\eps)$-dimensional framework, ${\cal D}^{(0,2)}_{313}$
could be considered as a new integral bases.

\item  From the result in \cite{Gluza:2010ws},  ${\cal A}^{(0,2)}_{313}$
can be written as linear combinations of integral bases ${\cal
A}^{(0,0)}_{313}$ and ${\cal A}^{(0,1)}_{313}$. However, the
coefficients depend on $\eps$. Then  $\eps \Delta {\cal
A}^{(0,0)}_{313}$ and $\eps \Delta {\cal A}^{(0,1)}_{313}$ could
contribute to finite terms, such as $\ln(-\chi)$, under the
unitarity cut.

\item In fact, $\ln\Big(\frac{-2\chi}{\gamma_i-1}\Big)$ is the result
given by unitarity cut channel $K_{12}$ of one-loop massless box
$(K_1, K_2, K_3, K_4)$ up to zero-order of $(\gamma_i-1)$. It
may indicate some connection with one-loop box diagram.

\end{itemize}

Finally for $b=3$ we found
\begin{eqnarray}
{\cal D}^{(0,3)}_{313}&=&\chi^2\ s^2{\cal D}^{(0,1)}_{313}
+\Big(\frac{5}{2}\chi^2+\chi-\frac{1}{4}\Big){\cal D}^{(0,0)}_{202}
-s\Big(\frac{3}{2}\chi^2+\frac{1}{2}\chi-\frac{1}{4}\Big)\Big({\cal
D}^{(0,0)}_{212}+{\cal D}^{(0,0)}_{302}\Big)\nn & &
-3\chi^2\ln(-\chi)~.~~~\label{b=3-result}
\end{eqnarray}
It is obvious that ${\cal D}^{(0,3)}_{313}$ can be written as linear
combination of ${\cal D}^{(0,i)}_{313}$, $i=0,1,2$(as well as lower
topologies) with rational functions of $\chi,s$.

\section{Conclusion}

In this paper we applied the unitarity method to two-loop diagrams
to determine their integral bases.  Two propagators for each loop
are cut while  mixed propagators are untouched. Integrations for the
reduced phase space have been done in the spinor form analytically.
Based on these results, analytical structures have been identified
and integral bases have been determined.

To demonstrate, we  applied our method to investigate the double-box
topology and its daughters, with appropriate choice of cut momenta
and kinematic region. For the ${\cal A}_{212}$ topology, we found
that there is only one scalar basis for the pure 4D case, while for
general $(4-2\epsilon)$-dimension, if we use the dimensional shifted
bases, there are 20 scalar bases under good renormalizability
conditions. For the ${\cal A}_{213}$ topology, there is also only
one scalar basis for the pure 4D case, but for the
$(4-2\epsilon)$-dimension, scalar bases are not enough even
considering the dimensional shifted bases. We found that there are
48 dimensional-shifted integral bases for renoramalizable theories.
For the ${\cal A}_{313}$ topology, it is difficult to get an exact
expression for general $(4-2\epsilon)$-dimension case. Thus we only
considered a specific case ${\cal A}_{313}^{(0,b)}$ with $T_2=K_1$.
We presented results to the zeroth-order and found three bases for
general $(4-2\epsilon)$-dimension if we do not allow coefficients
depending on $\eps$.

Based on the method demonstrated in this paper, several possible
directions can be done in the future. Firstly, for the ${\cal
A}_{313}$ topology, the exact result for the specific case $a=0$ is
still missing. The general value of $a$ should also be considered.
Secondly, topologies discussed in this paper are not the most
general cases. The most general configurations are those that each
vertex has external momenta attached as well as massive propagators.
Results of these more general cases are necessary. Thirdly, to
obtain a complete set of integral bases, we need to investigate
other topologies classified in \cite{Feng:2012bm}. Finally, besides
determining the integral bases, the unitarity method is  also
powerful for finding rational coefficients of bases in the
reduction. We expect that, after the complete set of integral bases
being obtained, such method can be useful for practical two-loop
calculations\footnote{See also a very interesting new method
\cite{Abreu:2014cla} }.

~\\~\\

\appendix

\section{Some useful formulae\label{A}}

In this section, we present some useful formulae appearing in
various calculations in the  paper.

~\\
{\bf Total derivative: }

For the holomorphic anomaly method, it is important to write an
expression into the total derivative form. Here we list results for
two typical inputs:
\bea { \Spbb{\ell|d\ell} \Spbb{\eta|\ell}^n\over
\Spab{\ell|P|\ell}^{n+2}} = \Spbb{d\ell|\partial\ell} \left( {1\over
(n+1) \Spab{\ell|P|\eta}} { \Spbb{\eta|\ell}^{n+1} \over
\Spab{\ell|P|\ell}^{n+1}} \right) ~,~~~\label{Dre-1} \eea
and
 \bea {{\Spbb{\ell|d\ell} \over
\Spab{\ell|P|\ell}\Spab{\ell|Q|\ell}}= {1\over \Spaa{\ell|P Q|\ell}}
\Spbb{d\ell|\partial\ell}\ln \left( {\Spab{\ell|P|\ell}\over
\Spab{\ell|Q|\ell}}\right)}~.~~~\label{Der-2}\eea
~\\
{\bf Pole of $\Spaa{\ell|QK|\ell}$:}

In the calculation, we will meet pole of the form $\Spaa{\ell|Q
K|\ell}$ frequently. It contains two poles and we need to separate
them. If both $Q,K$ are massless we can write it as
$\Spaa{\ell|Q}\Spbb{Q|K}\Spaa{K|\ell}$. If at least one of them is
massive, for example $K$, we can construct two massless momenta as
$P_i=Q+x_i K$, $i=1,2$, where
\bea x_{1,2} & = & {-2Q\cdot K\pm \sqrt{\Delta}\over 2K^2}~~,~~
\Delta=(2Q\cdot K)^2-4 Q^2 K^2~.~~~\label{x1-x2}\eea
Using this we have
\bea \Spaa{\ell|Q K|\ell}=
{\Spaa{\ell|P_1}\Spbb{P_1|P_2}\Spaa{\ell|P_2}\over
(x_1-x_2)}~,~~~\label{QK-factor}\eea
and
\bea & &Q={x_1P_2-x_2P_1\over x_1-x_2}~~,~~K={P_1-P_2\over
x_1-x_2}~~,~~2P_1\cdot P_2={-\Delta\over K^2}~,~~~\nn &
&x_1x_2={Q^2\over K^2}~~,~~x_1+x_2={-2Q\cdot K\over
K^2}~~,~~x_1-x_2={\sqrt{\Delta}\over K^2}~.~~~ \eea

~\\
{\bf Residue of high order pole:}

Poles we met are often not single poles. To read out residues of
poles with high order we can do as follows. Using the expression
\bea \frac{1}{\Spaa{\ell~(\eta-\tau
s)}^n}=\frac{d^{n-1}}{d\tau^{n-1}}\left(\frac{1}{(n-1)!\Spaa{\ell~s}^{n-1}}
\frac{1}{\Spaa{\ell~(\eta-\tau s)}}\right)\Big|_{\tau\to 0}\eea
with arbitrary auxiliary spinor $\ket{s}$, the residue of function
 ${1\over
\Spaa{\ell~\eta}^n}{N(\ket{\ell},|\ell])\over D(\ket{\ell},|\ell])}$
is then given by
\bea
\frac{d^{n-1}}{d\tau^{n-1}}\left(\frac{1}{(n-1)!\Spaa{\eta~s}^{n-1}}
{N(\ket{\eta-\tau s},|\eta])\over D(\ket{\eta-\tau
s},|\eta])}\right)\Big|_{\tau\to 0}~.~~~\label{multi-pole}\eea
It is very important to emphasize that the $\bket{\ell}$ part has
been set to $\bket{\eta}$, while the $\ket{\ell}$ is replaced by
$(\ket{\eta}-\tau \ket{s})$ before taking the derivative.
~\\
{\bf Evaluation of $\Spab{P_1|R|P_2}\Spab{P_2|S|P_1}$:}

We often encounter expression $\Spab{P_1|R|P_2}\Spab{P_2|S|P_1}$,
which  can be evaluated as
\bea \Spab{P_1|R|P_2}\Spab{P_2|S|P_1}&=&{\rm
tr}\left({1-\gamma_5\over
2}\not{P_1}\not{R}\not{P_2}\not{S}\right)\nn &=&2(P_1\cdot
R)(P_2\cdot S)+2(P_1\cdot S)(P_2\cdot R)-2(P_1\cdot P_2)(R\cdot
S)\nn & &-2i\epsilon(P_1RP_2S)~,~~~\eea
where $\epsilon(P_1RP_2S)$ denotes
$\epsilon_{\mu\nu\rho\sigma}P_1^\mu R^\nu P_2^\rho S^\sigma$. To
 evaluate $\epsilon(P_1RP_2S)^2$, a simple way is to  consider $\Spab{P_1|R|P_2}\Spab{P_2|S|P_1}
 \Spab{P_1|S|P_2}\Spab{P_2|R|P_1}$.

\section{Standard one-loop integrations\label{B}}

In this section we  list some standard one-loop results. We focus on
the following standard integral \cite{Britto:2006fc}
\bea  {\cal R}^{(a)}_{n}&\equiv   & \int
\Spaa{\la|d\la}\Spbb{\W\la|d\W\la} { \Spab{\la|R|\W\la}^a \over
\Spab{\la|K|\W\la}^{a+4-n}\prod_{i=1}^{n-2}\Spab{\la|Q_i|\W\la}}~,~~~
\label{R-stand-gen} \eea
which is the integration in \eref{1loop-uni-gen}. In our
application, we only need cases $n=2,3,4$.

\subsection{The bubble integration}

When $n=2$, we have $\Delta {\cal
A}^{(a)}_{2}=  \int d^{-2\eps} \mu [(1-2z) K^2]^{a+1}{\cal
R}^{(a)}_{2}$ with
\bea  {\cal R}^{(a)}_{2}&= &  \int
\Spaa{\la|d\la}\Spbb{\W\la|d\W\la} { \Spab{\la|R|\W\la}^a \over
\Spab{\la|K|\W\la}^{a+2}}\nn
& = & \int \Spaa{\la|d\la}\Spbb{d\W\la|{\d\over \partial\W\la}}
{1\over (a+1)\Spaa{\la|R K |\la}}{\Spab{\la|R|\W\la}^{a+1}  \over
\Spab{\la|K|\W\la}^{a+1}}~,~~~ \label{1loop-uni-gen-n=2} \eea
where \eref{Dre-1} has been used. For the pole $\Spaa{\la|R K
|\la}$, we use the construction given in Appendix \ref{A} to read
out two poles $\Spaa{\la|P_1}$ and $\Spaa{\la|P_2}$ with $P_i =  R+
x_i K$(see \eref{x1-x2}). For the first pole $\ket{\la}=\ket{P_1}$,
the residue is
\bean  {(x_1-x_2)\over (a+1)(- 2P_1\cdot P_2)}
(-x_1)^{a+1}~,~~~\eean
while for the second pole $\ket{\la}=\ket{P_2}$, the residue is
\bean {(x_1-x_2)\over (a+1)( 2P_1\cdot P_2)} (-x_2)^{a+1}~.~~~\eean
Putting them together we obtain
\bea {\cal R}^{(a)}_{2}& = & {1\over (a+1) \sqrt{\Delta_{R,K}}} (
(-x_1)^{a+1}-(-x_2)^{a+1})~,~~~\nn
\Delta {\cal A}^{(a)}_{2}& = & \int d^{-2\eps} \mu [(1-2z)
K^2]^{a+1}{\cal R}^{(a)}_{2}~,~~~\label{R-bubble}\eea
where
\bean \Delta_{R,K} =  (2R\cdot K)^2-4 R^2 K^2~~,~~x_1  =  {-2R\cdot
K+ \sqrt{\Delta_{R,K}}\over 2K^2}~~,~~x_2 = {-2R\cdot K-
\sqrt{\Delta_{R,K}}\over 2K^2}~.~~~ \eean
Let us give a few examples:
\bea \Delta {\cal A}^{(a=0)}_{2}& = & \int d^{-2\eps}\mu
(-\sqrt{1-u})~,~~~ \nn
\Delta {\cal A}^{(a=1)}_{2}& = & \int d^{-2\eps}\mu (-\sqrt{1-u})\{
K\cdot T\}~,~~~ \nn
\Delta {\cal A}^{(a=2)}_{2}& = & \int d^{-2\eps} \mu(-\sqrt{1-u})\{
{(4-u) (K\cdot T)^2 +(-1+u) K^2 T^2\over
3}\}~.~~~\label{1loop-n=2-a012}\eea
The case $a=0$ gives the analytic signature ${\cal
S}_{bub}=(-\sqrt{1-u})$ for the one-loop scalar bubble basis. For
cases $a=1,2$, the part inside the curly bracket is indeed
polynomial of $u$.

\subsection{The triangle integration}

For the case $n=3$, we can split the integrand as follows
\bea  {\cal R}^{(a)}_{3}&=  &  \int
\Spaa{\la|d\la}\Spbb{\W\la|d\W\la} { \Spab{\la|R|\W\la}^a \over
\Spab{\la|K|\W\la}^{a+1}\Spab{\la|Q|\W\la}} \nn
& = & \int \Spaa{\la|d\la}\Spbb{\W\la|d\W\la} \left\{
\left({\Spaa{\la|R Q|\la}\over \Spaa{\la| K Q|\la}} \right)^a
{1\over\Spab{\la|K|\W\la}\Spab{\la|Q|\W\la}}\right.\nn
& & +\left.\sum_{i=0}^{a-1} { \Spaa{\la|R K|\la}\over \Spaa{\la|Q
K|\la}} \left( {\Spaa{\la|R Q|\la}\over \Spaa{\la|K Q|\la}}\right)^i
{\Spab{\la|R|\W\la}^{a-1-i} \over
\Spab{\la|K|\W\la}^{a+1-i}}\right\}~.~~~
\label{1loop-uni-gen-n=3}\eea
After the splitting, the first term inside the big bracket  produces
the signature of triangle, while the second term produces the
signature of bubble. Thus we have the following two standard
integrations.
\subsubsection{Triangle-to-triangle part}

For the first term, writing into total derivative we have
\bea {\cal R}^{(a)}_{3\to 3} & = & \int \Spaa{\la|d\la}
\Spbb{d\W\la|{\d\over \d \W\la}} {(-)^a\Spaa{\la|R Q|\la}^a\over
\Spaa{\la|Q K|\la}^{a+1}} \ln\left({\Spab{\la|Q|\W\la}\over
\Spab{\la|K|\W\la}}\right)~.~~~\label{Stan-Tri-form}\eea
The pole is given by factor $\Spaa{\la|Q K|\la}^{a+1}$. Using
results in Appendix \ref{A}, for the pole $\eta= P_1$ with auxiliary
spinor  $s=P_2$ the residue is
\bean
 {(-)^a (x_1-x_2)^{a+1} \over
\Spbb{P_1|P_2}^{a+1}}\ln\left(-x_1\right) {d^a \over d\tau^a} \left(
{\Spaa{P_1-\tau P_2|R Q|P_1-\tau P_2}^a\over
a!\Spaa{P_1|P_2}^{2a+1}}\right)\Big|_{\tau\to 0}~.~~~ \eean
For the pole $\eta= P_2$ with auxiliary spinor  $s=P_1$  the residue
is
\bean {(-)^a (x_1-x_2)^{a+1} \over
\Spbb{P_1|P_2}^{a+1}}\ln\left(-x_2\right) {d^a \over d\tau^a} \left(
{\Spaa{P_2-\tau P_1|R Q|P_2-\tau P_1}^a\over
a!\Spaa{P_2|P_1}^{2a+1}}\right)\Big|_{\tau\to 0}~.~~~ \eean
One can observe that the derivative part is in fact the same for
both contributions after taking the limit $\tau\to 0$. Thus the sum
of two contributions is
\bean {(-)^a (x_1-x_2)^{a+1}\ln { x_1\over x_2} \over
\Spbb{P_1|P_2}^{a+1}a!\Spaa{P_1|P_2}^{2a+1}}
 {d^a \over d\tau^a}
\Spaa{P_1-\tau P_2|R Q|P_1-\tau P_2}^a\Big|_{\tau\to 0}~.~~~\eean

After some manipulation, we finally have
\bea {\cal R}^{(a)}_{3\to 3} & = & {\cal C}_{3\to 3}^{(a)}~ {\cal
S}_{tri}~,~~~\label{R3-to3}\eea
where ${\cal S}_{tri}$ is the signature of triangle and ${\cal
C}_{3\to 3}^{(a)}$ is the corresponding coefficient:
\bea {\cal S}_{tri}&\equiv&{1\over \sqrt{\Delta_{Q,K}}}\ln\left({ Q
\cdot K- \sqrt{  (Q\cdot K)^2 -Q^2 K^2}\over  Q \cdot K+ \sqrt{
(Q\cdot K)^2 -Q^2 K^2}}\right)~,~~~\nn {\cal C}_{3\to 3}^{(a)} & = &
{(-)^a\over a! \Delta_{Q,K}^a}
 {d^a\over d\tau^a} \left( +\tau ( 4 Q^2 (R\cdot K)-4 (R\cdot
Q)(Q\cdot K))+\tau^2 (Q^2) \right. \nn & & \left. + ( R^2
\Delta_{Q,K}+ (2 R\cdot Q)^2 K^2 + (2R\cdot K)^2 Q^2 - { (2R\cdot
Q)(2R\cdot K) (2Q\cdot K) })  \right)^a|_{\tau\to
0}~.~~~\label{C3-to3}\eea
The $a=0$ case gives the result for standard scalar triangle and
other $a$'s, give the corresponding coefficients under the
reduction. One can verify that the coefficients are indeed rational
functions.

\subsubsection{Triangle-to-bubble part}

The typical term in \eref{1loop-uni-gen-n=3} for triangle-to-bubble
part  is
\bea
{\cal R}_{3\to 2}[i,n] & \equiv &\int
\Spaa{\la|d\la}\Spbb{\W\la|d\W\la} { \Spaa{\la|R K|\la}\over
\Spaa{\la|Q K|\la}} \left( {\Spaa{\la|R Q|\la}\over \Spaa{\la|K
Q|\la}}\right)^i {\Spab{\la|R|\W\la}^n \over
\Spab{\la|K|\W\la}^{n+2}}\nn
& = & \int \Spaa{\la|d\la}\Spbb{d\W\la|{\d\over \d \W\la}} {(-)^i
\Spaa{\la|R Q|\la}^i\over (n+1)\Spaa{\la|Q K|\la}^{i+1}}
{\Spab{\la|R|\W\la}^{n+1} \over
\Spab{\la|K|\W\la}^{n+1}}~.~~~\label{Stan-3-to-2-form} \eea
The residue of pole $\Spaa{\la|Q K|\la}^{i+1}$ can be read out as in
previous subsubsection and we get
\bea {\cal R}_{3\to 2}[i,n] & = & { (-)^{n+i} (K^2)^i \over
i!(n+1)\sqrt{\Delta}^{n+2i+2}}{d^i\over d\tau^i}\left\{\left((2
R\cdot P_2 -\tau\Spab{P_1|R|P_2})^{n+1}\right.
 \right.\nn & & \left.
(-x_2 \Spab{P_2|R|P_1}-x_1 \tau^2\Spab{P_1|R|P_2} +\tau ( x_2
(2R\cdot P_1) +x_1(2R\cdot P_2)))^i\right)\nn & & +(-)^n \left((2
R\cdot P_1-\tau\Spab{P_2|R|P_1})^{n+1}\right. \nn & & \left.\left.
(-x_2\tau^2\Spab{P_2|R|P_1} -x_1 \Spab{P_1|R|P_2} +\tau ( x_2
(2R\cdot P_1) +x_1(2R\cdot P_2)))^i\right)\right\}\Big|_{\tau\to
0}~.~~~\label{R-3-to-2} \eea

To get a Lorentz contracted form, we need to use the following key
fact: {\sl to have non-zero contribution, factors $\Spab{P_1|R|P_2}$
and $\Spab{P_2|R|P_1}$ should always appear in pair}. Thus we can
transfer \eref{R-3-to-2} to
\bea {\cal R}_{3\to 2}[i,n] & = & { (-)^{n+i} (K^2)^i \over
i!(n+1)\sqrt{\Delta}^{n+2i+2}}{d^i\over d\tau^i}\left\{\left((2
R\cdot P_2 -\tau)^{n+1}\right.
 \right.\nn & & \left.
(-x_2 \Spab{P_2|R|P_1}\Spab{P_1|R|P_2}-x_1 \tau^2 +\tau ( x_2
(2R\cdot P_1) +x_1(2R\cdot P_2)))^i\right)\nn & & +(-)^n \left((2
R\cdot P_1-\tau)^{n+1}\right. \nn & & \left.\left. (-x_2\tau^2 -x_1
\Spab{P_1|R|P_2}\Spab{P_2|R|P_1} +\tau ( x_2 (2R\cdot P_1)
+x_1(2R\cdot P_2)))^i\right)\right\}\Big|_{\tau\to
0}~,~~~\label{R-3-to-3} \eea
where
\bea & &  \Spab{P_2|R|P_1} \Spab{P_1|R|P_2}=  {R^2 \Delta\over K^2}
+(2 R\cdot Q)^2  + (2R\cdot K)^2 {Q^2\over K^2} - { (2R\cdot
Q)(2R\cdot K) (2Q\cdot K) \over K^2}~.~~~\eea
Thus the contribution for the triangle-to-bubble part is given by
\bea {\cal R}_{3\to 2}^{(a)}= \sum_{i=0}^{a-1}{\cal R}_{3\to
2}[i,a-1-i]~.~~~\label{R3-to2-tot}\eea
Putting two parts together, we get
\bea {\cal R}_{3}^{(a)}={\cal R}_{3\to 3}^{(a)}+{\cal R}_{3\to
2}^{(a)}~.~~~\label{R3-tot}\eea
%

\subsection{The box integration}

The box integration is given by
\bea {\cal R}_4^{(a)}& = & \int \Spaa{\la|d\la}
\Spbb{\W\la|d\W\la}{\Spab{\la|R|\W\la}^a\over
\Spab{\la|K|\W\la}^{a}\Spab{\la|Q_1|\W\la}\Spab{\la|Q_2|\W\la}}~.~~~\label{1loo-R4}\eea
After splitting, we have the part producing signatures of box and
triangle
\bea &  & \int \Spaa{\la|d\la} \Spbb{\W\la|d\W\la}
\left\{{-\Spaa{\la|R Q_1|\la}\over \Spaa{\la|Q_1 Q_2|\la}} \left(
{\Spaa{\la|R Q_1|\la}\over \Spaa{\la| K Q_1|\la}} \right)^{a-1}
{1\over\Spab{\la|K|\W\la}\Spab{\la|Q_1|\W\la}}\right.\nn
& & \left. + {\Spaa{\la|R Q_2|\la}\over \Spaa{\la|Q_1 Q_2|\la}}
\left( {\Spaa{\la|R Q_2|\la}\over \Spaa{\la| K Q_2|\la}}
\right)^{a-1}
{1\over\Spab{\la|K|\W\la}\Spab{\la|Q_2|\W\la}}\right\}~,~~~\label{1loop-R4-1}\eea
and the part producing the signature of bubble
\bea &  & \int \Spaa{\la|d\la} \Spbb{\W\la|d\W\la}
\left\{{\Spaa{\la|R Q_2|\la}\over \Spaa{\la|Q_1
Q_2|\la}}\sum_{i=0}^{a-2} { \Spaa{\la|R K|\la}\over \Spaa{\la|Q_2
K|\la}} \left( {\Spaa{\la|R Q_2|\la}\over \Spaa{\la|K
Q_2|\la}}\right)^i {\Spab{\la|R|\W\la}^{a-2-i} \over
\Spab{\la|K|\W\la}^{a-i}}\right.\nn
& &\left. +{-\Spaa{\la|R Q_1|\la}\over \Spaa{\la|Q_1
Q_2|\la}}\sum_{i=0}^{a-2} { \Spaa{\la|R K|\la}\over \Spaa{\la|Q_1
K|\la}} \left( {\Spaa{\la|R Q_1|\la}\over \Spaa{\la|K
Q_1|\la}}\right)^i {\Spab{\la|R|\W\la}^{a-2-i} \over
\Spab{\la|K|\W\la}^{a-i}}\right\}~.~~~\label{1loop-R4-2}\eea
Now we can evaluate various parts one by one.

\subsubsection{The box-to-box part}

This part comes from pole  $\Spaa{\la|Q_1 Q_2|\la}$ in
\eref{1loop-R4-1}. Using $Q_1+x_i Q_2$ to construct two null momenta
$P_i$, we get the residue
\bean &  & {(x_1-x_2)\over \Spbb{P_1|P_2} \Spaa{P_1|P_2}} \left(
{\Spab{P_1|R|P_2}\over \Spab{P_1|K|P_2}}\right)^a\ln (-x_1) +
{-(x_1-x_2)\over \Spbb{P_1|P_2} \Spaa{P_1|P_2}} \left(
{\Spab{P_2|R|P_1}\over \Spab{P_2|K|P_1}}\right)^a\ln (-x_2)~,~~~
\eean
which can be written as
\bea & & {1\over 2}{(x_1-x_2)\over \Spbb{P_1|P_2} \Spaa{P_1|P_2}}
\ln \left({ x_1\over x_2}\right) \left[  \left(
{\Spab{P_1|R|P_2}\over \Spab{P_1|K|P_2}}\right)^a+ \left(
{\Spab{P_2|R|P_1}\over \Spab{P_2|K|P_1}}\right)^a\right]\nn
& & +{1\over 2}{(x_1-x_2)\over \Spbb{P_1|P_2} \Spaa{P_1|P_2}} \ln
\left( x_1x_2\right) \left[  \left( {\Spab{P_1|R|P_2}\over
\Spab{P_1|K|P_2}}\right)^a- \left( {\Spab{P_2|R|P_1} \over
\Spab{P_2|K|P_1}}\right)^a\right]~.~~~\label{box-middle} \eea

~\\

{\bf Box part:} The first term in \eref{box-middle} produces the
signature of box
\bea {\cal S}_{box}={1\over \sqrt{ (2 Q_1\cdot Q_2)^2 -4 Q_1^2
Q_2^2}}\ln {Q_1\cdot Q_2- \sqrt{ ( Q_1\cdot Q_2)^2 - Q_1^2
Q_2^2}\over Q_1\cdot Q_2+ \sqrt{ ( Q_1\cdot Q_2)^2 - Q_1^2
Q_2^2}}~,~~~ \eea
as well as the coefficient
\bea {\cal C}_{4\to 4}^{(a)}= {1\over 2}\left[  \left(
{\Spab{P_1|R|P_2}\over \Spab{P_1|K|P_2}} \right)^a+ \left(
{\Spab{P_2|R|P_1}\over
\Spab{P_2|K|P_1}}\right)^a\right]~.~~~\label{Stan-box-box-coeff}\eea
Thus we have
\bea {\cal R}_{4\to 4}^{(a)}={\cal C}_{4\to 4}^{(a)}~ {\cal
S}_{box}~.~~~\label{Stan-box-box}\eea
It can be shown that there is a recursion relation
\bea {\cal R}_{4\to 4}^{(a+1)}={T_2\over T_1}{\cal R}_{4\to
4}^{(a)}-{T_3\over T_1}{\cal R}_{4\to 4}^{(a-1)}~,~~~\eea
with
\bea {\cal C}_{4\to 4}^{(0)} & = & 1~~,~~ {\cal C}_{4\to 4}^{(1)} =
{T_2\over 2 T_1}~,~~~\eea
where
\bean T_1 & = &  4 [ (Q_1\cdot K)^2+ { Q_1^2\over Q_2^2} (Q_2\cdot
K)^2- {2 Q_1\cdot Q_2\over Q_2^2} (Q_1\cdot K)(Q_2\cdot K)] + K^2
{((2 Q_1\cdot Q_2)^2 -4 Q_1^2 Q_2^2)\over Q_2^2}~,~~~\nn
T_2 & = & { 8 (R\cdot K)(( Q_1\cdot Q_2)^2 - Q_1^2 Q_2^2)\over
Q_2^2} + 8 (R\cdot Q_1) (K\cdot Q_1)+ 8 (R\cdot Q_2) (K\cdot
Q_2){Q_1^2 \over Q_2^2}\nn & & -8 { (Q_1\cdot Q_2)\over Q_2^2} (
(R\cdot Q_1) (K\cdot Q_2)+(R\cdot Q_2) (K\cdot Q_1))~,~~~\nn
 T_3 & = &
 4 [ (Q_1\cdot R)^2+ { Q_1^2\over
Q_2^2} (Q_2\cdot R)^2- {2 Q_1\cdot Q_2\over Q_2^2} (Q_1\cdot
R)(Q_2\cdot R)] + R^2 {((2 Q_1\cdot Q_2)^2 -4 Q_1^2 Q_2^2)\over
Q_2^2}~.~~~ \eean

~\\

{\bf Triangle part:} The second term in \eref{box-middle} produces
the signature of triangle. Using
\bean \ln (x_1x_2)=\ln {Q_1^2\over Q_2^2}= \ln {Q_1^2\over K^2}- \ln
{Q_2^2\over K^2}~,~~~ \eean
the second term in \eref{box-middle} can be rewritten as
\bea & &  {1\over 2}\ln {Q_1^2\over K^2}\left\{{1\over \Spaa{\la|Q_1
Q_2|\la}} \left( {\Spaa{\la|R Q_1|\la}\over \Spaa{\la| K Q_1|\la}}
\right)^{a}\right\}_{{\rm Residue~of}~\Spaa{\la|Q_1 Q_2|\la}}\nn
& & + {1\over 2}\ln {Q_2^2\over K^2}\left\{ {-1\over \Spaa{\la|Q_1
Q_2|\la}} \left( {\Spaa{\la|R Q_2|\la}\over \Spaa{\la| K Q_2|\la}}
\right)^{a}\right\}_{{\rm Residue~of}~\Spaa{\la|Q_1
Q_2|\la}}~.~~~\label{4-to-3-tripart}\eea
We will combine \eref{4-to-3-tripart} with results in the next
subsubsection to produce the complete triangle part.

\subsubsection{The box-to-triangle part}

Since $Q_1$ and $Q_2$ are symmetric, we will focus on the triangle
constructed by $K,Q_1$. The contribution comes from the first term
of \eref{1loop-R4-1}. This term contains two kinds of poles:
$\Spaa{\la|Q_1 Q_2|\la}$ and  $\Spaa{\la|K Q|\la}$. The contribution
of pole  $\Spaa{\la|Q_1 Q_2|\la}$ has been evaluated in previous
subsubsection. For the second pole, after writing it into total
derivative, it is ${1\over \Spaa{\la|K Q|\la}^a}$. Using $Q_1, K$ to
construct two null momenta $P_1, P_2$,  the residue is given by two
parts. The first part contains $\ln(x_1 x_2)$ (which is nothing but
$\ln {Q_1^2\over K^2}$) and is given by
\bea {1\over 2}\ln {Q_1^2\over K^2}\left\{{1\over \Spaa{\la|Q_1
Q_2|\la}} \left( {\Spaa{\la|R Q_1|\la}\over \Spaa{\la| K Q_1|\la}}
\right)^{a}\right\}_{{\rm Residue~of}~\Spaa{\la|K
Q_1|\la}^a}~.~~~\eea
It cancels the first term of \eref{4-to-3-tripart}, since the sum of
all residues of a holomorphic function is zero\footnote{It is worth
to notice that by power counting, infinity does not contribute
residue.}. The second part contains $\ln(x_1/x_2)$ which is the
signature of triangle. The contribution can be written as
\bea {\cal R}_{4\to 3}^{(a)}(Q_1) = {\cal C}_{4\to 3}^{(a)}(Q_1)~
{\cal S}_{tri}(Q_1, K)~,~~~\label{Stan-box-tri} \eea
where
\bean {\cal S}_{tri}(Q_1, K)= {1\over \sqrt{ (2 Q_1\cdot K)^2 -4
Q_1^2 K^2}}\ln {Q_1\cdot K- \sqrt{ ( Q_1\cdot K)^2 - Q_1^2 K^2}\over
Q_1\cdot K+ \sqrt{ ( Q_1\cdot K)^2 - Q_1^2 K^2}}~,~~~\eean
and
\bea {\cal C}_{4\to 3}^{(a)}(Q_1) & = & { (-)^{a-1}\over (a-1)!}
\left( {K^2\over 4((Q_1\cdot K)^2-K^2
Q_1^2)}\right)^{a-1}~~~\label{C-box-tri-1} \\ & & \left\{
{d^{a-1}\over d\tau^{a-1}}\left( {(-\tau^2 x_2 \Spab{P_2|R|P_1} -x_1
\Spab{P_1|R|P_2} +\tau ( x_2
\Spab{P_1|R|P_1}+x_1\Spab{P_2|R|P_2}))^a \over (-\tau^2 x_2
\Spab{P_2|Q_2|P_1} -x_1 \Spab{P_1|Q_2|P_2} +\tau ( x_2
\Spab{P_1|Q_2|P_1}+x_1\Spab{P_2|Q_2|P_2}))}\right)\right.\nn
& & + \left. {d^{a-1}\over d\tau^{a-1}}\left( {(- x_2
\Spab{P_2|R|P_1} -x_1\tau^2 \Spab{P_1|R|P_2} +\tau ( x_2
\Spab{P_1|R|P_1}+x_1\Spab{P_2|R|P_2}))^a \over (- x_2
\Spab{P_2|Q_2|P_1} -x_1\tau^2 \Spab{P_1|Q_2|P_2} +\tau ( x_2
\Spab{P_1|Q_2|P_1}+x_1\Spab{P_2|Q_2|P_2}))}\right)\right\}\Big|_{\tau\to
0}~.~~~\nonumber\eea
To write the spinor form to the Lorentz contracted form, we can take
similar manipulation as the one from \eref{R-3-to-2} to
\eref{R-3-to-3}. The result is
\bea {\cal C}_{4\to 3}^{(a)}(Q_1) & = &{(-)^{a-1}\over
(a-1)!}\left({K^2\over 4((Q_1\cdot K)^2-Q_1^2K^2)}\right)^{a-1}\nn &
& {d^{a-1}\over
d\tau^{a-1}}{2K^2T_6(K^2T_7+K^2T_5T_6\tau+Q_1^2T_6T_7\tau^2)\Big(1+{Q_1^2\over
K^2}T_6\tau^2+T_4\tau\Big)^a \over
(Q_1^2T_6T_7\tau^2+K^2(T_7+T_5T_6\tau))^2-T_8^2(K^2-Q_1^2T_6\tau^2)^2}\Big|_{\tau\to
0}~,~~~\label{C-box-tri-2} \eea
where we have defined
\bean T_4&=&{4(R\cdot K)Q_1^2-4(R\cdot Q_1)(K\cdot Q_1)\over
K^2}~~,~~T_5={4(Q_2\cdot K)Q_1^2-4(Q_2\cdot Q_1)(K\cdot Q_1)\over
K^2}~,~~~\nn T_6&=&{R^2\Delta\over K^2}+4(R\cdot
Q_1)^2+{4Q_1^2(R\cdot K)^2\over K^2}-{8(R\cdot Q_1)(R\cdot
K)(Q_1\cdot K)\over K^2}~,~~~\nn T_7&=&2(P_1\cdot Q_2)(P_2\cdot
R)+2(P_1\cdot R)(Q_2\cdot P_2)-2(P_1\cdot P_2)(Q_2\cdot R)~,~~~\nn
T_8&=&{4i\epsilon(Q_1Q_2K R)\sqrt{(K\cdot Q_1)^2-K^2Q_1^2}~\over
K^2}~.~~~ \eean
It is worth to mention that $T_8$ appears as $T_8^2$, thus the
Levi-Civita symbol has been removed.

\subsubsection{The box-to-bubble part}

Having finished the computation of \eref{1loop-R4-1},  we turn to
the \eref{1loop-R4-2}.   The total result can be expressed as
\bea {\cal R}_{4\to 2}^{(a)}=\sum_{i=0}^{a-2}{\cal R}_{4\to
2}(Q_1)[i,a-1-i]+\left\{Q_1\leftrightarrow Q_2\right\}~,~~~\eea
where the typical term is
\bea {\cal R}_{4\to 2}(Q_1) [i,m]=\left\{{1\over
\Spaa{\la|Q_1Q_2|\la}}\left({\Spaa{\la|RQ_1|\la}\over
\Spaa{\la|KQ_1|\la}}\right)^{i+1}{1\over
m}{\Spab{\la|R|\W\la}^m\over\Spab{\la|K|\W\la}^m}\right\}_{residue}~.~~~
\eea
There are three poles for this part:  $\Spaa{\la|Q_1Q_2|\la}$,
$\Spaa{\la|Q_1K|\la}$ and $\Spaa{\la|Q_2K|\la}$. The contribution
from $\Spaa{\la|Q_1Q_2|\la}$ is zero when summing up the two lines
in \eref{1loop-R4-2}. For the remaining two poles, because of the
symmetry $Q_1\leftrightarrow Q_2$, we will focus on ${\cal R}_{4\to
2}(Q_1)[i,m]$ only.

We use $Q_1,~K$ to construct two null momenta $P_1,~P_2$ and get
residue
\bea & & {\cal R}_{4\to2}(Q_1)[i,m]\nn &=&{(-)^i(K^2)^i\over
i!m(\sqrt{\Delta})^{m+2i+1}}{d^i\over d\tau^i}\left\{{(\tau
T_4-\tau^2x_2\Spab{P_2|R|P_1}-x_1\Spab{P_1|R|P_2})^{i+1}(\Spab{P_1|R|P_1}-\tau\Spab{P_2|R|P_1})^m\over
\tau T_5-\tau^2
x_2\Spab{P_2|Q_2|P_1}-x_1\Spab{P_1|Q_2|P_2}}\right.\nn &
&\left.+(-)^{m+1}{(\tau
T_4-x_2\Spab{P_2|R|P_1}-\tau^2x_1\Spab{P_1|R|P_2})^{i+1}(\Spab{P_2|R|P_2}-\tau\Spab{P_1|R|P_2})^m\over
\tau T_5-
x_2\Spab{P_2|Q_2|P_1}-\tau^2x_1\Spab{P_1|Q_2|P_2}}\right\}\Big|_{\tau\to
0}~.~~~\label{R-4-2-Q1ia} \eea
We can rewrite the expression to the following Lorentz contracted
form
\bea & &{\cal R}_{4\to 2}(Q_1)[i,a]={(-)^i(K^2)^i\over
i!a(\sqrt{\Delta})^{a+2i+1}}{d^i\over d\tau^i}\left\{{(\tau
T_4+\tau^2{Q_1^2\over K^2}T_6+1)^{i+1} (2Q_1\cdot R+2x_1K\cdot
R+\tau x_1T_6)^a\over \tau T_5+\tau^2{Q_1^2\over
K^2}(T_7-T_8)+{T_7+T_8\over T_6}}\right.\nn &
&\left.+(-)^{a+1}{(\tau T_4+\tau^2{Q_1^2\over K^2}T_6+1)^{i+1}
(2Q_1\cdot R+2x_2K\cdot R+\tau x_2T_6)^a\over \tau
T_5+\tau^2{Q_1^2\over K^2}(T_7+T_8)+{T_7-T_8\over T_6}}\right\}~.~~~
\eea
One can verify that $T_8$ will appear as $T_8^2$ after summing
${\cal R}_{4\to 2}(Q_1)[i,a]$ and ${\cal R}_{4\to 2}(Q_2)[i,a]$,
thus the Levi-Civita symbol does not appear in the final result.

\section{The integration for topology ${\cal A}_{313}$  }

It is hard to get the explicit result for \eref{A-exp-3}. In this
Appendix we develop a method to find approximate expressions.
Technically the case $b=0$ is the most complicated one, while the
$b\geq 1$ cases can be reduced to the case $b=0$ plus some simple
integration. Before working out the integration case by case, we
give two explicit integrations
\begin{eqnarray}\label{integrate left}
\int^{+1}_{-1}\frac{du}{\sqrt{(u+m\gamma_2)^2+(1-m^2)(\gamma^2_2-1)}}=\ln\Big(\frac{\gamma_2+1}{\gamma_2-1}\Big)~,~~~
\end{eqnarray}
and
\begin{eqnarray}\label{integrate right}
\int^{+1}_{-1}\frac{du}{\sqrt{(u+\alpha)^2-\beta^2}}\ln\frac{(u+\alpha)+\sqrt{(u+\alpha)^2-\beta^2}}{(u+\alpha)-\sqrt{(u+\alpha)^2-\beta^2}}
=\ln\Big(\frac{\gamma+1}{\gamma-1}\Big)\ln\Big(\frac{\gamma_1+1}{\gamma_1-1}\Big)~,~~~
\end{eqnarray}
where we have used the conditions $-1<m<1$ and $\gamma_i\geq1$.
These two results are useful for our further discussion.

\subsection{Pure 4D solution in the case $b=0$}

From (\ref{new expression a=0}) we get that ${\cal D}^{(0,0)}_{313}$
is
\begin{eqnarray}
-\frac{\gamma_1\gamma_2}{s^3}\int^{+1}_{-1}du
\frac{1}{\sqrt{(u+m\gamma_2)^2+\xi^2}}
\frac{1}{\sqrt{(u+\alpha)^2-\beta^2}}\ln\frac{(u+\alpha)+\sqrt{(u+\alpha)^2-\beta^2}}{(u+\alpha)-\sqrt{(u+\alpha)^2-\beta^2}}~,~~~
\end{eqnarray}
where $\xi^2= (1-m^2)(\gamma^2_2-1)$ is positive. In the pure 4D,
$\xi\rightarrow0$, so we need to study the limit behavior at
$\xi\rightarrow0$. If we expand
\begin{eqnarray}\label{def f(u)}
f(u)\equiv\frac{1}{\sqrt{(u+\alpha)^2-\beta^2}}\ln\frac{(u+\alpha)+\sqrt{(u+\alpha)^2-\beta^2}}{(u+\alpha)-\sqrt{(u+\alpha)^2-\beta^2}}
=\sum^{+\infty}_{n=0}f_n u^n
\end{eqnarray}
in the region $[-1,+1]$,  where  $f(u)$ is positive and convergent
uniformly,  we will have(ignoring the factor
$(-\gamma_1\gamma_2/s^3)$)
\begin{eqnarray}\label{span A313_00}
{\cal
D}^{(0,0)}_{313}=\sum^{+\infty}_{n=0}f_n\sum^{n}_{k=0}C^k_n(-m\gamma_2)^{n-k}
\int^{m\gamma_2+1}_{m\gamma_2-1}du\frac{u^k}{\sqrt{u^2+\xi^2}}
\end{eqnarray}
after shifting of $u$. Now we introduce a series of functions
defined as
\begin{eqnarray}
H_n(a,b)=\int^b_0dx\ \frac{x^n}{\sqrt{x^2+a^2}}~~,~~n\geq 0~,~~~\end{eqnarray}
with integer $n$. It is easy to figure out the answers
\begin{eqnarray}
&&
H_{n=2m}=\frac{(-)^{m}a^n}{2^n}C^m_n\ln\Big(\frac{\sqrt{a^2+b^2}+b}{a}\Big)+\frac{a^n}{2^n}
\sum^{m-1}_{k=0}\frac{(-)^{k}C^{k}_{n}}{n-2k}\Bigg[
\Big(\frac{\sqrt{a^2+b^2}+b}{a}\Big)^{n-2k}-\Big(\frac{\sqrt{a^2+b}-b}{a}\Big)^{n-2k}\Bigg]~,~~~
\nonumber \\
&&
H_{n=2m+1}=\frac{a^n}{2\sqrt{\pi}}\Gamma\Big(-\frac{n}{2}\Big)\Gamma\Big(\frac{n+1}{2}\Big)
+\frac{a^n}{2^n}\sum^m_{k=0}\frac{(-)^{k}C^k_n}{n-2k}\Bigg[\Big(\frac{\sqrt{a^2+b^2}+b}
{a}\Big)^{n-2k}+\Big(\frac{\sqrt{a^2+b^2}-b}{a}\Big)^{n-2k}\Bigg]~.~~~
\nonumber
\end{eqnarray}
For the limit $a\rightarrow 0$, it is easy to see that only in the
case $n=0$ it is  divergent and we have
\begin{eqnarray}\label{lim Hn}
\lim_{a\rightarrow 0}H_0(a,b)=\ln\Big(\frac{\sqrt{a^2+b^2}+b}{a}\Big)\Big|_{a\rightarrow0}~~,~~
\lim_{a\rightarrow 0}H_n(a,b)=\frac{b^n}{n}~~,~~n\geq 1~.~~~\end{eqnarray}
Using this observation the expression (\ref{span A313_00}) can be
separated into the divergent part and the finite part. The divergent
part is
\begin{eqnarray}
\sum^{+\infty}_{n=0}f_n(-m\gamma_2)^{n}\Big(H_0(\xi,m\gamma_2+1)-H_0(\xi,m\gamma_2-1)\Big)
=f(-m\gamma_2)\ln\Big(\frac{\gamma_2+1}{\gamma_2-1}\Big)~~~\label{A313
00 divergent}
\end{eqnarray}
by using the conditions $\gamma_2>1$ and $-1<m<1$, where function
$f$ is defined in (\ref{def f(u)}). Under the 4D limit, \textbf{the
divergent term} is
\begin{eqnarray}\label{divergent term lim}
{\cal D}^{(0,0)}_{313}|_{div}=\frac{1}{s^3}\
\frac{1}{2\chi}\ln\Big(\frac{\gamma_2+1}{\gamma_2-1}\Big)
\Bigg[2\ln(-\chi)+\ln\Big(\frac{\gamma+1}{\gamma-1}\Big)+\ln\Big(\frac{\gamma_1+1}{\gamma_1-1}\Big)\Bigg]~,~~~
\end{eqnarray}
where we have recovered the missing factor. If we consider ${\cal
D}^{(0,0)}_{313}$ as a series of $\gamma_2$, this is just the first
(divergent) term. The finite part of the expression (\ref{span
A313_00}) is
\begin{eqnarray}
\sum^{+\infty}_{n=0}f_n\sum^{n}_{k=1}C^k_n(-m\gamma_2)^{n-k}\Big(H_k(\xi,m\gamma_2+1)-H_k(\xi,m\gamma_2-1)\Big)~.~~~
\end{eqnarray}
Under the pure 4D limit, using (\ref{lim Hn}) it becomes
\begin{eqnarray}
\sum^{+\infty}_{n=0}f_n\sum^{n}_{k=1}\frac{C^k_n}{k}(-m\gamma_2)^{n-k}\Big((m\gamma_2+1)^k+(m\gamma_2-1)^k\Big)~.~~~
\end{eqnarray}
We can use parameterizing method to sum up above awesome form. If we
define
\begin{eqnarray}
G(x)\equiv\sum^{n}_{k=1}\frac{C^k_n}{k}(-m\gamma_2)^{n-k}\Big((m\gamma_2+x)^k+(m\gamma_2-x)^k\Big)~,~~~
\end{eqnarray}
then $G(x)$ satisfies the differential equation
\begin{eqnarray}
\frac{\partial G}{\partial x}=g(x)-g(-x)~~,~~g(x)=\frac{x^n-\rho^n}{x-\rho}~~,~~\rho\equiv-m\gamma_2~.~~~
\end{eqnarray}
Obviously,
\begin{eqnarray}\label{G(1)-G(0)}
G(1)-G(0)=\int^{+1}_0dx\ g(x)+\int^{-1}_0dx\ g(x)~,~~~
\end{eqnarray}
where $G(1)$ is the result we want to find. To compute $G(0)$, we
define new function
\begin{eqnarray}
\W
  G(0,x)=2\sum^{n}_{k=1}\frac{C^k_n}{k}(-m\gamma_2)^{n-k}(m\gamma_2\
x)^k~~,~~\W G(0,1)=G(0),~\W G(0,0)=0~.~~~
\end{eqnarray}
Using the same method, we can find the differential equation for $\W
G(0,x)$. After some variable replacement we get
\begin{eqnarray}\label{G(0)}
G(0)=\W G(0,1)=2\int^0_{\rho}g(x)\ dx~.~~~
\end{eqnarray}
Combining (\ref{G(1)-G(0)}) with (\ref{G(0)}) and exchanging the
integration and the summation
\begin{eqnarray}
\sum^{+\infty}_{n=0}f_n\int dx\ \frac{x^n-\rho^n}{x-\rho}=\int dx\
\frac{f(x)-f(\rho)}{x-\rho}~,~~~
\end{eqnarray}
finally \textbf{the finite term} can be written as
\begin{eqnarray}\label{finite term integral}
{\cal
D}^{(0,0)}_{313}|_{finite}=\frac{-\gamma_1\gamma_2}{s^3}\Big(\int^{+1}_\rho+\int^{-1}_\rho\Big)dx\
\frac{f(x)-f(\rho)}{x-\rho}~.~~~
\end{eqnarray}

Now we focus on the indefinite integration. If we change the
integration variable as
\begin{eqnarray}\label{int variable replacment}
\cosh y\equiv\frac{\alpha+x}{\beta}~~,~~\cosh
y_\pm\equiv\frac{\alpha\pm 1}{\beta}~~,~~\cosh
y_0\equiv\frac{\alpha-m\gamma_2}{\beta}~,~~~
\end{eqnarray}
%
then
\begin{eqnarray}
\int dx\ \frac{f(x)}{x-\rho}=\frac{2}{\beta}\int dy\ \frac{y}{\cosh
y-\cosh y_0}~.~~~
\end{eqnarray}
After integration by parts it becomes
\begin{eqnarray}
\frac{2}{\beta\sinh
y_0}\Bigg[-\frac{y^2}{2}+y\ln\frac{1-e^{-(y_0-y)}}{1-e^{-(y_0+y)}}
+\textsf{Li}_2\Big(e^{-(y_0-y)}\Big)+\textsf{Li}_2\Big(e^{-(y_0+y)}\Big)\Bigg]~,~~~
\end{eqnarray}
in which $\textsf{Li}_s(z)$ is the polylogarithm. Combining with the
other part, the whole indefinite integral of (\ref{finite term
integral}) can be written as
\begin{eqnarray}
F(y)&\equiv&\frac{2}{\beta\sinh y_0}\Bigg[\textsf{Li}_2\Big(e^{-(y_0-y)}\Big)+\textsf{Li}_2\Big(e^{-(y_0+y)}\Big)-\frac{y^2}{2} \nonumber \\
&&-(y_0-y)\ln\Big(1-e^{-(y_0-y)}\Big)-(y_0+y)\ln\Big(1-e^{-(y_0+y)}\Big)\Bigg]~.~~~
\end{eqnarray}
Thus ${\cal D}^{(0,0)}_{313}|_{finite}$ is given by
$F(y_+)+F(y_-)-2F(y_0)$ up to an overall factor. Above calculations
are done for pure 4D limit of $\gamma_2$. After taking the pure 4D
limit of $\gamma_1,\gamma$ we finally reach
\begin{eqnarray}
{\cal D}^{(0,0)}_{313}=\frac{1}{s^3\ 2\chi}\Bigg[
&&\ln\Big(\frac{-2\chi}{\gamma-1}\Big)\ln\Big(\frac{-2\chi}{\gamma_1-1}\Big)+\ln\Big(\frac{-2\chi}{\gamma-1}\Big)\ln\Big(\frac{-2\chi}{\gamma_2-1}\Big)
+\ln\Big(\frac{-2\chi}{\gamma_1-1}\Big)\ln\Big(\frac{-2\chi}{\gamma_2-1}\Big) \nonumber \\
&&+2\ \textsf{Li}_2(1+\chi)-\frac{\pi^2}{3}\Bigg]~,~~~
\end{eqnarray}
after combining with the divergent term (\ref{divergent term lim}).

\subsection{Pure 4D solution in the case $b\geq 1$}

{\bf For the case b=1} we can define a combination of $(b=1)$ and
$(b=0)$ as
\begin{eqnarray}
{\cal D}^{(0,1)}_{313}-\chi\ s{\cal
D}^{(0,0)}_{313}=\frac{\gamma_1}{2s^2}\int^{+1}_{-1}du
\frac{u+m\gamma_2}{\sqrt{(u+m\gamma_2)^2+\xi^2}}\ f(u)~,~~~
\end{eqnarray}
and again $f(u)$ defined in (\ref{def f(u)}). Since
\begin{eqnarray}
\Big|\frac{u+m\gamma_2}{\sqrt{(u+m\gamma_2)^2+\xi^2}}\Big|\ \leq1
\end{eqnarray}
and $f(u)>0$ in the whole integration zone, using the result
(\ref{integrate right}) we have
\begin{eqnarray}
{\cal D}^{(0,1)}_{313}-\chi\ s{\cal D}^{(0,0)}_{313}\ <\
\frac{\gamma_1}{2s^2}\ln\Big(\frac{\gamma+1}{\gamma-1}\Big)\ln\Big(\frac{\gamma_1+1}{\gamma_1-1}\Big)~.~~~
\end{eqnarray}
It means that as a function of $\gamma_2$, above combination is
finite under the limit $\gamma_2\rightarrow1$. Thus to our
zero-order(i.e., the pure 4D case), we can just set $\gamma_2=1$
before doing the integration and get
\begin{eqnarray}
{\cal D}^{(0,1)}_{313}-\chi\ s{\cal
D}^{(0,0)}_{313}=\frac{\gamma_1}{2s^2}\int^{+1}_{-1}du\frac{u+m}{|u+m|}f(u)
=\frac{\gamma_1}{2s^2}\Big(\int^{+1}_{-m}+\int^{-1}_{-m}\Big)du\
f(u)~.~~~
\end{eqnarray}
Taking the same integration variable replacement as in (\ref{int
variable replacment}), this integral can be worked out easily. Then
in the limit $\gamma_2\rightarrow1$, ${\cal D}^{(0,1)}_{313}-\chi\
s{\cal D}^{(0,0)}_{313}$ is equal to
\begin{eqnarray}\label{A313 01 r2=1}
\frac{\gamma_1}{2s^2}\Bigg[\ln\frac{\gamma\gamma_1-m+\alpha_m}{(\gamma-1)(\gamma_1-1)}\ln\frac{\gamma\gamma_1-m-\alpha_m}{(\gamma-1)(\gamma_1-1)}
+\ln\frac{\gamma\gamma_1-m+\alpha_m}{(\gamma+1)(\gamma_1-1)}\ln\frac{\gamma\gamma_1-m-\alpha_m}{(\gamma+1)(\gamma_1-1)}\Bigg]~,~~~
\end{eqnarray}
where
\begin{eqnarray}
\alpha_m=\sqrt{\gamma^2+\gamma^2_1+m^2-2\gamma\gamma_1m-1}~.~~~
\end{eqnarray}
It is worth to point out that when we take $m\rightarrow1$, this
result reduces to
\begin{eqnarray}
\frac{\gamma_1}{2s^2}\ln\Big(\frac{\gamma+1}{\gamma-1}\Big)\ln\Big(\frac{\gamma_1+1}{\gamma_1-1}\Big)~,~~~
\end{eqnarray}
which is just the explicit result of ${\cal D}^{(0,0)}_{312}$. To
keep only zero-order results, we take \
$\gamma,\gamma_1\rightarrow1$\ further in (\ref{A313 01 r2=1}) and
find that  in the pure 4D
\begin{eqnarray}
{\cal D}^{(0,1)}_{313}=\chi\ s{\cal D}^{(0,0)}_{313}+{\cal
D}^{(0,0)}_{312}
-\frac{1}{s^2}\ln\Big(\frac{-2\chi}{\gamma-1}\Big)\ln\Big(\frac{-2\chi}{\gamma_1-1}\Big)~.~~~
\end{eqnarray}

{\bf For the case b=2,3} we will not show the computation details
again. The main point is that in the first step, we choose a proper
linear combination of ${\cal D}^{(0,b)}_{313}$ to make the integrand
having the form
\begin{eqnarray}
\frac{(u+m\gamma_2)^b}{\sqrt{(u+m\gamma_2)^2+\xi^2}}\ f(u)~.~~~
\end{eqnarray}
For $b=2$ we should choose the combination as
\begin{eqnarray}
{\cal D}^{(0,2)}_{313}-2\chi\ s{\cal D}^{(0,1)}_{313}+\chi^2\
s^2{\cal D}^{(0,0)}_{313}~,~~~
\end{eqnarray}
and for $b=3$ it is
\begin{eqnarray}
{\cal D}^{(0,3)}_{313}-3\chi\ s{\cal D}^{(0,2)}_{313}+3\chi^2\
s^2{\cal D}^{(0,1)}_{313}-\chi^3\ s^3{\cal D}^{(0,0)}_{313}~.~~~
\end{eqnarray}
Then we can prove that these combinations are convergent at
$\gamma_2\rightarrow1$ just as  in the case $b=1$. Thus we can take
$\gamma_2=1$ before integrating those combinations. The second step
is to integrate these dramatically simplified integrands. In this
step the most efficient way is to use the variable replacement in
(\ref{int variable replacment}) and we can integrate quickly. For
example in the case $b=2$, combination \ ${\cal
D}^{(0,2)}_{313}-\chi\ s{\cal D}^{(0,1)}_{313}$ is given by
\begin{eqnarray}
-\frac{\gamma_1}{2s}\Bigg(\gamma\ln\Big(\frac{\gamma_1+1}{\gamma_1-1}\Big)+\gamma_1\ln\Big(\frac{\gamma+1}{\gamma-1}\Big)
-2\alpha_m\ln\frac{\gamma\gamma_1-m+\alpha_m}{\beta}-2m\Bigg)~,~~~
\end{eqnarray}
and the corresponding expressions for $b=3$ are even longer. To find
the approximate results in the pure 4D,  in the last step we take
limit $\gamma,\gamma_1\rightarrow1$. Carrying out these steps,
finally we get
\begin{eqnarray}\label{result b=2,3}
{\cal D}^{(0,2)}_{313}&=&\chi\ s{\cal
D}^{(0,1)}_{313}+\frac{2\chi+1}{s}{\cal D}^{(0,0)}_{202}
-\frac{2\chi+1}{2}{\cal D}^{(0,0)}_{212}-\frac{2\chi+1}{2}{\cal D}^{(0,0)}_{302}-\frac{2\chi}{s}\ln(-\chi)~,~~~\\
{\cal D}^{(0,3)}_{313}&=&\chi^2\ s^2{\cal D}^{(0,1)}_{313}
+\Big(\frac{5}{2}\chi^2+\chi-\frac{1}{4}\Big){\cal D}^{(0,0)}_{202}
-\Big(\frac{3}{2}\chi^2+\frac{1}{2}\chi-\frac{1}{4}\Big)\Big(s{\cal
D}^{(0,0)}_{212}+s{\cal D}^{(0,0)}_{302}\Big)-3\chi^2\ln(-\chi)~.~~~
\nonumber
\end{eqnarray}

\section*{Acknowledgement}

We would like to thank  Mingxing Luo for early participant of the
project and  Ruth Britto, David Kosower, Song He, Yang
Zhang for valuable discussions. We would also like to thank all
organizers and  participants of "Amplitudes 2013" from April 28 to
May 3, 2013 in Tegernsee of Germany, where part of results has been
presented. R.H thanks Niels Bohr International Academy, Niels Bohr
Institute in Denmark for the supporting during his PhD study. R.H's
research is supported by the European Research Council under
Advanced Investigator Grant ERC-AdG-228301. B.F, K.Z and J.Z are
supported, in part, by fund from Qiu-Shi and Chinese NSF funding
under contracts No.11031005, No.11135006, No.11125523.



\begin{thebibliography}{References}


\bibitem{Bern:2008ef}
  Z.~Bern {\it et al.}  [NLO Multileg Working Group],
  arXiv:0803.0494 [hep-ph].

\bibitem{Binoth:2010ra}
  J.~R.~Andersen {\it et al.}  [SM and NLO Multileg Working Group Collaboration],
   arXiv:1003.1241 [hep-ph].  

\bibitem{AlcarazMaestre:2012vp}
  J.~Alcaraz Maestre {\it et al.}  [SM AND NLO MULTILEG and SM MC Working Groups Collaboration],
  arXiv:1203.6803 [hep-ph].  



\bibitem{Berger:1983yi}
  E.~L.~Berger, E.~Braaten and R.~D.~Field,
  Nucl.\ Phys.\ B {\bf 239}, 52 (1984).

\bibitem{Aurenche:1985yk}
  P.~Aurenche, A.~Douiri, R.~Baier, M.~Fontannaz and D.~Schiff,
  Z.\ Phys.\ C {\bf 29}, 459 (1985).

\bibitem{Ellis:1987xu}
  R.~K.~Ellis, I.~Hinchliffe, M.~Soldate and J.~J.~van der Bij,
  Nucl.\ Phys.\ B {\bf 297}, 221 (1988).



\bibitem{Smirnov:2004ym}
  V.~A.~Smirnov,
  ``Evaluating Feynman integrals,''  Springer Tracts Mod.\ Phys.\  {\bf 211}, 1 (2004).

\bibitem{Smirnov:2006ry}
  V.~A.~Smirnov,
  ``Feynman integral calculus,''  Berlin, Germany: Springer (2006) 283 p

\bibitem{Smirnov:2012gma}
  V.~A.~Smirnov,
  ``Analytic tools for Feynman integrals,''  Springer Tracts Mod.\ Phys.\  {\bf 250}, 1 (2012).



\bibitem{Chetyrkin:1981qh}
  K.~G.~Chetyrkin and F.~V.~Tkachov,
  Nucl.\ Phys.\ B {\bf 192}, 159 (1981).

\bibitem{Tarasov:1998nx}
  O.~V.~Tarasov,
  Acta Phys.\ Polon.\ B {\bf 29}, 2655 (1998)  [hep-ph/9812250].

\bibitem{Bern:2000dn}
  Z.~Bern, L.~J.~Dixon and D.~A.~Kosower,
  JHEP {\bf 0001}, 027 (2000)  [hep-ph/0001001].

\bibitem{Anastasiou:2000kg}
  C.~Anastasiou, E.~W.~N.~Glover, C.~Oleari and M.~E.~Tejeda-Yeomans,
  Nucl.\ Phys.\ B {\bf 601}, 318 (2001)  [hep-ph/0010212].

\bibitem{Anastasiou:2000ue}
  C.~Anastasiou, E.~W.~N.~Glover, C.~Oleari and M.~E.~Tejeda-Yeomans,
  Nucl.\ Phys.\ B {\bf 601}, 341 (2001)  [hep-ph/0011094].

\bibitem{Glover:2001af}
  E.~W.~N.~Glover, C.~Oleari and M.~E.~Tejeda-Yeomans,
  Nucl.\ Phys.\ B {\bf 605}, 467 (2001)  [hep-ph/0102201].

\bibitem{Anastasiou:2001sv}
  C.~Anastasiou, E.~W.~N.~Glover, C.~Oleari and M.~E.~Tejeda-Yeomans,
  Nucl.\ Phys.\ B {\bf 605}, 486 (2001)  [hep-ph/0101304].

\bibitem{Laporta:2001dd}
  S.~Laporta,
  Int.\ J.\ Mod.\ Phys.\ A {\bf 15}, 5087 (2000)  [hep-ph/0102033].

\bibitem{Bern:2002tk}
  Z.~Bern, A.~De Freitas and L.~J.~Dixon,
  JHEP {\bf 0203}, 018 (2002)  [hep-ph/0201161].

\bibitem{Tarasov:2004ks}
  O.~V.~Tarasov,
  Nucl.\ Instrum.\ Meth.\ A {\bf 534}, 293 (2004)  [hep-ph/0403253].



\bibitem{Gluza:2010ws}
  J.~Gluza, K.~Kajda and D.~A.~Kosower,
   Phys.\ Rev.\ D {\bf 83}, 045012 (2011)  [arXiv:1009.0472 [hep-th]].  

\bibitem{Kalmykov:2011yy}
  M.~Y.~.Kalmykov and B.~A.~Kniehl,
   Phys.\ Lett.\ B {\bf 702}, 268 (2011)  [arXiv:1105.5319 [math-ph]].  

\bibitem{Schabinger:2011dz}
  R.~M.~Schabinger,
  JHEP {\bf 1201}, 077 (2012)  [arXiv:1111.4220 [hep-ph]].  





\bibitem{Kotikov:1990kg}
  A.~V.~Kotikov,
  Phys.\ Lett.\ B {\bf 254}, 158 (1991).

\bibitem{Remiddi:1997ny}
  E.~Remiddi,
  Nuovo Cim.\ A {\bf 110}, 1435 (1997)  [hep-th/9711188].

\bibitem{Gehrmann:1999as}
  T.~Gehrmann and E.~Remiddi,
  Nucl.\ Phys.\ B {\bf 580}, 485 (2000)  [hep-ph/9912329].

\bibitem{Argeri:2007up}
  M.~Argeri and P.~Mastrolia,
  Int.\ J.\ Mod.\ Phys.\ A {\bf 22}, 4375 (2007)  [arXiv:0707.4037 [hep-ph]].

\bibitem{Henn:2013pwa}
  J.~M.~Henn,
  Phys.\ Rev.\ Lett.\  {\bf 110}, 251601 (2013)  [arXiv:1304.1806 [hep-th]].

\bibitem{Henn:2013woa}
  J.~M.~Henn and V.~A.~Smirnov,
  JHEP {\bf 1311}, 041 (2013)  [arXiv:1307.4083].

\bibitem{Henn:2013nsa}
  J.~M.~Henn, A.~V.~Smirnov and V.~A.~Smirnov,
  arXiv:1312.2588 [hep-th].

\bibitem{Argeri:2014qva}
  M.~Argeri, S.~Di Vita, P.~Mastrolia, E.~Mirabella, J.~Schlenk, U.~Schubert and L.~Tancredi,
  arXiv:1401.2979 [hep-ph].



\bibitem{Bergere:1973fq}
  M.~C.~Bergere and Y.~-M.~P.~Lam,
  Commun.\ Math.\ Phys.\  {\bf 39}, 1 (1974).

\bibitem{Usyukina:1975yg}
  N.~I.~Usyukina,
  Teor.\ Mat.\ Fiz.\  {\bf 22}, 300 (1975).

\bibitem{Smirnov:1999gc}
  V.~A.~Smirnov,
  Phys.\ Lett.\ B {\bf 460}, 397 (1999)  [hep-ph/9905323].

\bibitem{Tausk:1999vh}
  J.~B.~Tausk,
  Phys.\ Lett.\ B {\bf 469}, 225 (1999)  [hep-ph/9909506].



\bibitem{Passarino:1978jh}
  G.~Passarino and M.~J.~G.~Veltman,
  Nucl.\ Phys.\ B {\bf 160}, 151 (1979).

\bibitem{Neerven:1984}
  W.L van Neerven and J.A.M Vermaseren,
  "Large loop integrals,"
  Phys.\ Lett.\ B {\bf 137}B, 241(1984).

\bibitem{Bern:1992}
   Z.~Bern, L.~Dixon and D.A.~Kosower,
   Phys.\ Lett.\ B {\bf 302}, 299(1993) [ERRATUM-ibid.\ B {\bf 318}, 649(1993)]
   [arXiv:hep-ph/9212308];
   Z.~Bern, L.~Dixon and D.A.~Kosower,
   Phys.\ Phys.\ B {\bf 412}, 751(1994) [arXiv:hep-ph/9306240].

\bibitem{Ellis:2007}
   R.~K.~Ellis and G.~Zanderighi,
   JHEP {\bf 0802}, 002(2008) [arXiv:0712.1851 [hep-ph]];
   A.~Denner and S.~Dittmaier,
   Nucl.\ Phys.\ B {\bf734}, 62(2006) [arXiv:hep-ph/0509141];
   G.~Duplancic, B.~Nizic,
   Eur.\ Phys.\ J.\ C {\bf 35}, 105 (2004) [arXiv:hep-ph/0303184].



\bibitem{Bern:1994zx}
  Z.~Bern, L.~J.~Dixon, D.~C.~Dunbar and D.~A.~Kosower,
  Nucl.\ Phys.\ B {\bf 425}, 217 (1994)  [hep-ph/9403226];
  Z.~Bern, L.~J.~Dixon, D.~C.~Dunbar and D.~A.~Kosower,
  Nucl.\ Phys.\ B {\bf 435}, 59 (1995)  [hep-ph/9409265];
  Z.~Bern, L.~J.~Dixon and D.~A.~Kosower,
  Ann.\ Rev.\ Nucl.\ Part.\ Sci.\  {\bf 46}, 109 (1996)  [hep-ph/9602280].

\bibitem{Bern:1995db}
  Z.~Bern and A.~G.~Morgan,
  Nucl.\ Phys.\ B {\bf 467}, 479 (1996)  [hep-ph/9511336].

\bibitem{Bern:1997sc}
  Z.~Bern, L.~J.~Dixon and D.~A.~Kosower,
  Nucl.\ Phys.\ B {\bf 513}, 3 (1998)  [hep-ph/9708239].

\bibitem{Bern:1996ja}
  Z.~Bern, L.~J.~Dixon, D.~C.~Dunbar and D.~A.~Kosower,
  Phys.\ Lett.\ B {\bf 394}, 105 (1997)  [hep-th/9611127].

\bibitem{Britto:2004nc}
  R.~Britto, F.~Cachazo and B.~Feng,
  Nucl.\ Phys.\ B {\bf 725}, 275 (2005)  [hep-th/0412103].

\bibitem{Britto:2004nj}
  R.~Britto, F.~Cachazo and B.~Feng,
  Phys.\ Rev.\ D {\bf 71}, 025012 (2005)  [hep-th/0410179];
  S.~J.~Bidder, N.~E.~J.~Bjerrum-Bohr, L.~J.~Dixon and D.~C.~Dunbar,
  Phys.\ Lett.\ B {\bf 606}, 189 (2005)  [hep-th/0410296];
  S.~J.~Bidder, N.~E.~J.~Bjerrum-Bohr, D.~C.~Dunbar and W.~B.~Perkins,
  Phys.\ Lett.\ B {\bf 612}, 75 (2005)  [hep-th/0502028];
  S.~J.~Bidder, D.~C.~Dunbar and W.~B.~Perkins,
  JHEP {\bf 0508}, 055 (2005)  [hep-th/0505249];
  Z.~Bern, N.~E.~J.~Bjerrum-Bohr, D.~C.~Dunbar and H.~Ita,
  JHEP {\bf 0511}, 027 (2005)  [hep-ph/0507019].

\bibitem{Bern:2005cq}
  Z.~Bern, L.~J.~Dixon and D.~A.~Kosower,
  Phys.\ Rev.\ D {\bf 73}, 065013 (2006)  [hep-ph/0507005].

\bibitem{Britto:2005ha}
  R.~Britto, E.~Buchbinder, F.~Cachazo and B.~Feng,
  Phys.\ Rev.\ D {\bf 72}, 065012 (2005)  [hep-ph/0503132];
  R.~Britto, B.~Feng and P.~Mastrolia,
  Phys.\ Rev.\ D {\bf 73}, 105004 (2006)  [hep-ph/0602178];
   P.~Mastrolia,
  Phys.\ Lett.\ B {\bf 644}, 272 (2007)  [hep-th/0611091].

\bibitem{Brandhuber:2005jw}
  A.~Brandhuber, S.~McNamara, B.~J.~Spence and G.~Travaglini,
  JHEP {\bf 0510}, 011 (2005)  [hep-th/0506068].

\bibitem{Bern:2007dw}
  Z.~Bern, L.~J.~Dixon and D.~A.~Kosower,
  Annals Phys.\  {\bf 322}, 1587 (2007)  [arXiv:0704.2798 [hep-ph]].

\bibitem{Forde:2007mi}
  D.~Forde,
  Phys.\ Rev.\ D {\bf 75}, 125019 (2007)  [arXiv:0704.1835 [hep-ph]].

\bibitem{Badger:2008cm}
  S.~D.~Badger,
  JHEP {\bf 0901}, 049 (2009)  [arXiv:0806.4600 [hep-ph]].

\bibitem{Anastasiou:2006jv}
  C.~Anastasiou, R.~Britto, B.~Feng, Z.~Kunszt and P.~Mastrolia,
  Phys.\ Lett.\ B {\bf 645}, 213 (2007)  [hep-ph/0609191];
  C.~Anastasiou, R.~Britto, B.~Feng, Z.~Kunszt and P.~Mastrolia,
  JHEP {\bf 0703}, 111 (2007)  [hep-ph/0612277];
  W.~T.~Giele, Z.~Kunszt and K.~Melnikov,
  JHEP {\bf 0804}, 049 (2008)  [arXiv:0801.2237 [hep-ph]].

\bibitem{Britto:2006fc}
  R.~Britto and B.~Feng,
  Phys.\ Rev.\ D {\bf 75}, 105006 (2007)  [hep-ph/0612089];
  R.~Britto and B.~Feng,
  JHEP {\bf 0802}, 095 (2008)  [arXiv:0711.4284 [hep-ph]];
  R.~Britto, B.~Feng and P.~Mastrolia,
  Phys.\ Rev.\ D {\bf 78}, 025031 (2008)  [arXiv:0803.1989 [hep-ph]];
  R.~Britto, B.~Feng and G.~Yang,
  JHEP {\bf 0809}, 089 (2008)  [arXiv:0803.3147 [hep-ph]];
  B.~Feng and G.~Yang,
   Nucl.\ Phys.\ B {\bf 811}, 305 (2009)  [arXiv:0806.4016 [hep-ph]];
  R.~Britto and B.~Feng,
  Phys.\ Lett.\ B {\bf 681}, 376 (2009)  [arXiv:0904.2766 [hep-th]].

\bibitem{Berger:2009zb}
  C.~F.~Berger and D.~Forde,
  Ann.\ Rev.\ Nucl.\ Part.\ Sci.\  {\bf 60}, 181 (2010)  [arXiv:0912.3534 [hep-ph]].

\bibitem{Bern:2010qa}
  Z.~Bern, J.~J.~Carrasco, T.~Dennen, Y.~-t.~Huang and H.~Ita,
  Phys.\ Rev.\ D {\bf 83}, 085022 (2011)  [arXiv:1010.0494 [hep-th]].



\bibitem{Britto:2010xq}
  R.~Britto,
  J.\ Phys.\ A {\bf 44}, 454006 (2011)  [arXiv:1012.4493 [hep-th]].

\bibitem{Dixon:2013uaa}
  L.~J.~Dixon,
  arXiv:1310.5353 [hep-ph].  




\bibitem{Buchbinder:2005wp}
  E.~I.~Buchbinder and F.~Cachazo,
  JHEP {\bf 0511}, 036 (2005)  [hep-th/0506126].  
  F.~Cachazo,
   arXiv:0803.1988 [hep-th].  
  F.~Cachazo, M.~Spradlin and A.~Volovich,
   Phys.\ Rev.\ D {\bf 78}, 105022 (2008)  [arXiv:0805.4832 [hep-th]].  


\bibitem{ArkaniHamed:2009dn}
  N.~Arkani-Hamed, F.~Cachazo, C.~Cheung and J.~Kaplan,
   JHEP {\bf 1003}, 020 (2010)  [arXiv:0907.5418 [hep-th]].  


\bibitem{ArkaniHamed:2012nw}
  N.~Arkani-Hamed, J.~L.~Bourjaily, F.~Cachazo, A.~B.~Goncharov, A.~Postnikov and J.~Trnka,



\bibitem{Kosower:2011ty}
  D.~A.~Kosower and K.~J.~Larsen,
  Phys.\ Rev.\ D {\bf 85}, 045017 (2012)  [arXiv:1108.1180 [hep-th]].

\bibitem{Larsen:2012sx}
  K.~J.~Larsen,
  Phys.\ Rev.\ D {\bf 86}, 085032 (2012)  [arXiv:1205.0297 [hep-th]].

\bibitem{CaronHuot:2012ab}
  S.~Caron-Huot and K.~J.~Larsen,
  JHEP {\bf 1210}, 026 (2012)  [arXiv:1205.0801 [hep-ph]].

\bibitem{Johansson:2012zv}
  H.~Johansson, D.~A.~Kosower and K.~J.~Larsen,
  Phys.\ Rev.\ D {\bf 87}, 025030 (2013)  [arXiv:1208.1754 [hep-th]].

\bibitem{Johansson:2012sf}
  H.~Johansson, D.~A.~Kosower and K.~J.~Larsen,
  PoS LL {\bf 2012}, 066 (2012)  [arXiv:1212.2132].

\bibitem{Sogaard:2013yga}
  M.~Sogaard,
  JHEP {\bf 1309}, 116 (2013)  [arXiv:1306.1496 [hep-th]].

\bibitem{Johansson:2013sda}
  H.~Johansson, D.~A.~Kosower and K.~J.~Larsen,
  arXiv:1308.4632 [hep-th].

\bibitem{Sogaard:2013fpa}
  M.~Sogaard and Y.~Zhang,
  JHEP {\bf 1312}, 008 (2013)  [arXiv:1310.6006 [hep-th]].



\bibitem{Ossola:2006us}
  G.~Ossola, C.~G.~Papadopoulos and R.~Pittau,
  Nucl.\ Phys.\ B {\bf 763}, 147 (2007)  [hep-ph/0609007].



\bibitem{Mastrolia:2011pr}
  P.~Mastrolia and G.~Ossola,
  JHEP {\bf 1111}, 014 (2011)  [arXiv:1107.6041 [hep-ph]].

\bibitem{Badger:2012dp}
  S.~Badger, H.~Frellesvig and Y.~Zhang,
  JHEP {\bf 1204}, 055 (2012)  [arXiv:1202.2019 [hep-ph]].

\bibitem{Mastrolia:2012an}
  P.~Mastrolia, E.~Mirabella, G.~Ossola and T.~Peraro,
  Phys.\ Lett.\ B {\bf 718}, 173 (2012)  [arXiv:1205.7087 [hep-ph]].

\bibitem{Kleiss:2012yv}
  R.~H.~P.~Kleiss, I.~Malamos, C.~G.~Papadopoulos and R.~Verheyen,
  JHEP {\bf 1212}, 038 (2012)  [arXiv:1206.4180 [hep-ph]].

\bibitem{Badger:2012dv}
  S.~Badger, H.~Frellesvig and Y.~Zhang,
  HEP {\bf 1208}, 065 (2012)  [arXiv:1207.2976 [hep-ph]].

\bibitem{Mastrolia:2012wf}
  P.~Mastrolia, E.~Mirabella, G.~Ossola and T.~Peraro,
  Phys.\ Rev.\ D {\bf 87}, no. 8, 085026 (2013)  [arXiv:1209.4319 [hep-ph]].

\bibitem{Huang:2013kh}
  R.~Huang and Y.~Zhang,
  JHEP {\bf 1304}, 080 (2013)  [arXiv:1302.1023 [hep-ph]].

\bibitem{Badger:2013gxa}
  S.~Badger, H.~Frellesvig and Y.~Zhang,
  JHEP {\bf 1312}, 045 (2013)  [arXiv:1310.1051 [hep-ph]].

\bibitem{Zhang:2012ce}
  Y.~Zhang,
  JHEP {\bf 1209}, 042 (2012)  [arXiv:1205.5707 [hep-ph]].

\bibitem{Feng:2012bm}
  B.~Feng and R.~Huang,
   JHEP {\bf 1302}, 117 (2013)  [arXiv:1209.3747 [hep-ph]].

\bibitem{Mastrolia:2012du}
  P.~Mastrolia, E.~Mirabella, G.~Ossola, T.~Peraro and H.~van Deurzen,
  PoS LL {\bf 2012}, 028 (2012)  [arXiv:1209.5678 [hep-ph]].




\bibitem{Cachazo:2004kj}
  F.~Cachazo, P.~Svrcek and E.~Witten,
  JHEP {\bf 0409}, 006 (2004)  [hep-th/0403047].


\bibitem{Heinrich:2010ax}
  G.~Heinrich, G.~Ossola, T.~Reiter and F.~Tramontano,
  JHEP {\bf 1010}, 105 (2010)  [arXiv:1008.2441 [hep-ph]].

\bibitem{Abreu:2014cla}
  S.~Abreu, R.~Britto, C.~Duhr and E.~Gardi,
  arXiv:1401.3546 [hep-th].  



\end{thebibliography}
\end{document}